\documentclass[aps,prc,twocolumn,showpacs]{revtex4}

\usepackage[T1]{fontenc}
\usepackage[latin1]{inputenc}
\usepackage[english]{babel}

\usepackage{textcomp}
\usepackage{amsmath}
\usepackage{amsfonts}
\usepackage{amssymb}
\usepackage{units}

\usepackage{graphicx}
\usepackage{xcolor}

\usepackage{ifpdf}
\ifpdf
  \newcommand*{\hyperrefcolor}{violet}
  \usepackage[%
    pdfsubject={Manuscript},%
    pdftitle={Manuscript},%
    pdfauthor={us},%
    bookmarksopen=true,   
    plainpages=false,     
    colorlinks,           
    linkcolor=\hyperrefcolor,%
    anchorcolor=\hyperrefcolor,%
    citecolor=\hyperrefcolor,%
    urlcolor=\hyperrefcolor,%
    menucolor=\hyperrefcolor,%
    filecolor=\hyperrefcolor%
  ]{hyperref}
\fi

\def\level55L{5/2^+5/2[633](0.0)}

\begin{document}

\title{Radiative lifetime and energy of the low-energy isomeric level in
$^{229}$Th}

\author{E.~V.~Tkalya}
\email{eugene.tkalya@mail.ru}

\affiliation{Skobeltsyn Institute of Nuclear Physics Lomonosov Moscow State
University, Leninskie gory, Moscow 119991, Russia}
\affiliation{Nuclear Safety Institute of Russian Academy of Science, Bol'shaya
Tulskaya 52, Moscow 115191, Russia}

\author{Christian~Schneider}
\author{Justin~Jeet}
\author{Eric~R.~Hudson}

\affiliation{Department of Physics and Astronomy, University of California, Los
Angeles, California 90095, USA}

\date{\today}

\begin{abstract}
We estimate the range of the radiative lifetime and energy of the
anomalous, low-energy $3/2^+(7.8 \pm 0.5$~eV) state in the
$^{229}$Th nucleus. Our phenomenological calculations are based on the available
experimental data for the intensities of $M1$ and $E2$ transitions
between excited levels of the $^{229}$Th nucleus in the
$K^{\pi}[Nn_Z\Lambda]=5/2^+[633]$ and $3/2^+[631]$ rotational
bands. We also discuss the influence of certain branching
coefficients, which affect the currently accepted measured energy
of the isomeric state. From this work, we establish a favored
region where the transition lifetime and energy should lie at
roughly the 90\% confidence level. We also suggest new nuclear
physics measurements, which would significantly reduce the
ambiguity in the present data.
\end{abstract}

\maketitle

\section{Introduction}
\label{sec:Introduction}

The low-energy isomeric state in the $^{229}$Th nucleus is
currently a subject of intense experimental and theoretical
research (see a short review of the literature in \cite{Browne-08,
Jeet-15} and references below). This state is expected to provide
access to a number of interesting physical effects, including the
decay of the nuclear isomeric level via the electronic bridge
mechanism in certain chemical environments
\cite{Strizhov-91,Porsev-10_3+,Porsev-10_1+}, cooperative
spontaneous emission \cite{Dicke-54} in a system of excited
nuclei, the M\"{o}\ss{}bauer effect in the optical range
\cite{Tkalya-11}, sensitive tests of the variation of the fine
structure constant and the strong interaction parameter
\cite{Flambaum-06,He-07,Hayes-07,Litvinova2009}, a check of the exponentiality
of the decay law of an isolated metastable state at long times
\cite{Dykhne-98}, and accelerated $\alpha$-decay of the $^{229}$Th
nuclei via the low energy isomeric state \cite{Dykhne-96}. In
addition, two applications that may have a significant
technological impact were proposed: a new metrological standard
for time \cite{Tkalya-96} or the ``nuclear clock''
\cite{Peik-03,Rellergert-10,Campbell-12,Peik2015}, and a nuclear
laser (or gamma ray laser) in the optical range \cite{Tkalya-11}.

The ground state of the $^{229}$Th nucleus, $J^{\pi}E=5/2^+(0.0)$,
is the ground level of the rotational band
$K^{\pi}[Nn_Z\Lambda]=5/2^+[633]$. Currently, there is little
doubt in the existence of the low-energy isomeric level
$J^{\pi}E=3/2^+(E_{is})$, which is the lowest level of the
rotational band $K^{\pi}[Nn_Z\Lambda]=3/2^+[631]$. The existence
of the other levels of this band is reliably experimentally
validated \cite{Browne-08}. In addition, an independent
corroboration of the existence of a low-lying state has been
achieved experimentally in the reaction $^{230}$Th$(d,t)^{229}$Th
\cite{Burke-08}. This experiment provides strong evidence that the
$K^{\pi}[Nn_Z\Lambda]=3/2^+[631]$ band head is located very close
to the ground state, and, in fact, all available experimental data
from these indirect measurements of the
$J^{\pi}K[Nn_Z\Lambda]=3/2^+3/2[631]$  isomeric state energy
indicate that $E_{is}<10$~eV \cite{Helmer-94, Guimaraes-05,
Beck-07, Beck-09}.

Unfortunately, the energy resolution of such experiments does not
provide the accuracy required for direct optical spectroscopy of
the nuclear isomeric $M$1 transition
$5/2^+5/2[633](0.0)\leftrightarrow3/2^+3/2[631](E_{is})$, which is
clearly a prerequisite for the aforementioned studies. Therefore,
new approaches to determine the isomeric energy are required.
While there are some ongoing attempts to better measure the
isomeric transition energy (see for example \cite{Zhao-12}),
directly driving the nuclear transition inside an insulator with a
large band gap (i.e.{} a crystal), first proposed in the works
\cite{Tkalya-00-JETPL,Tkalya-00-PRC}, or in sample of trapped ions
\cite{Kaelber1989,Campbell-09,Herrera-Sancho-12} appear to be the
most promising routes forward in the short term.

In the crystal approach, a band gap greater than $E_{is}$ results
in the absence of the conversion decay channel of the low energy
isomeric state. Thus, the uncertainty in the decay probability,
which is associated with electronic conversion, disappears. In
Ref. \cite{Rellergert-10,Rellergert-10b,Hehlen-13} the
requirements and characteristics of the requisite crystals were
made rigorous, showing that $^{229}$Th:LiCaAlF$_6$ and
$^{229}$Th:LiSrAlF$_6$ were likely good choices; other efforts
focus on CaF$_2$ \cite{Dessovic-14,Stellmer2015}, or ThF$_4$ and
Na$_2$ThF$_6$ \cite{Ellis-14}. Several experiments using this
solid-state approach  have been carried out in recent years
\cite{Zhao-12,Yamaguchi-15,Jeet-15}, though as described in
Sec.~\ref{sec:Lifetime}, care must be taken to interpret the
results of these experiments.

Experiments using trapped Th$^+$
\cite{Herrera-Sancho-12,Okhapkin-15} or Th$^{3+}$
\cite{Campbell-09,Campbell-11,Radnaev-12,Beloy-14} ions aim at
exploiting the electronic bridge process
\cite{Strizhov-91,Porsev-10_3+,Porsev-10_1+}, which can dominate
the direct radiative decay of the isomeric transition. Using the
electronic bridge process and exciting the isomeric transition in
a multi-photon nucleon-electron simultaneous transition has the
potential advantage of obviating the use of a vacuum ultraviolet
laser system, at the expense of a modest increase in system
complexity. Experiments have advanced rapidly in recent years and
it is expected that with recent high-resolution electronic spectra
\cite{Campbell-11,Radnaev-12,HerreraSancho2013,Okhapkin-15},
electronic bridge excitation rates can be better calculated in the
near future.

In any experiment searching for the nuclear energy level, two key
parameters in the preparation of the experiment are the isomeric
state lifetime and energy. Therefore, the aim of this manuscript
is to provide a critical assessment and estimation of these
parameters to aid these experiments. In
Section~\ref{sec:MatrixElement} of this paper, we analyze the
experimental data on the nuclear matrix elements of transitions
between states belonging to rotational bands
$K^{\pi}[Nn_Z\Lambda]=5/2^+[633]$ and $3/2^+[631]$  both in
$^{229}$Th only and in comparable nuclei. In Section
\ref{sec:Lifetime}, we estimate the radiative lifetime of the
isomeric state using available experimental data for the
transition rates of the interband $M$1 and $E$2 gamma transitions
between excited levels of the $^{229}$Th nucleus. In
Section~\ref{sec:Conversion} we consider the importance of the
conversion decay channel, showing that these processes will
dominate the isomer radiative decay in cases where they are
possible and, therefore, must be avoided. In
Section~\ref{sec:BranchingCoefficients}, we analyze the possible
range of branching coefficients. We show how this range affects
the determination of the isomeric energy in the experiments of
Refs. \cite{Beck-07,Beck-R}, which provide the currently accepted
isomeric transition energy range. In Section
\ref{sec:FavoredArea}, we briefly discuss the results of this work
and present a ``favored region'', which we recommend the community
adopt in order to direct future searches. We conclude in
Section~\ref{sec:Results} with a summary of these results and
point out new nuclear physics measurements that should be
performed to considerably reduce the uncertainty in the present
data.

In the present work, we use (if not noted otherwise) the following
system of units: $\hbar=c=1$.

\section{Matrix element of the isomeric transition}
\label{sec:MatrixElement}

Together with the energy of the isomeric level, the magnitude of
the nuclear transition matrix element determines the half life
$T_{1/2}$ or the radiative lifetime $\tau$ ($\tau=T_{1/2}/\ln(2)$)
of the isomeric state. Currently, there are two possibilities for
phenomenological estimation of the reduced probability for the
isomeric transition, $B(M1;3/2^+3/2[631]\rightarrow
5/2^+5/2[633])$. The first possibility is to use parameters of the
similar 311.9 keV transition in the spectrum of the $^{233}$U
nucleus. The second is to take advantage of the available
experimental data for the $M$1 transitions between the rotational
bands $K^{\pi}[Nn_Z\Lambda]=5/2^+[633]$ and $3/2^+[631]$ in the
excitation spectrum of the $^{229}$Th nucleus.

The first method can be motivated, because transitions
$3/2^+3/2[631]\rightarrow 5/2^+5/2[633]$ at 311.9 keV in the
$^{233}$U nucleus and 7.8~eV in the $^{229}$Th nucleus look
identical in terms of the rotational model and, in that context,
should have the same reduced transition probabilities. (In this
and the following section, we will use the updated value of Ref.
\cite{Beck-R} for the $^{229}$Th isomeric energy, $E_{is}=7.8 \pm
0.5$~eV; see Section~\ref{sec:BranchingCoefficients} for further
discussion of the isomeric state energy.) The reduced probability
of the transition in the $^{233}$U nucleus is known to be
$B_{W.u.}(M1;3/2^+3/2[631](311.9$ keV$) \rightarrow
5/2^+5/2[633](0.0)) = (0.33\pm 0.05) \times 10^{-2}$
\cite{Singh-05}. Here, $B_{W.u.}$ denotes a reduced probability in
Weisskopf units \cite{Blatt-52} (see Appendix \ref{AppendixA} for
details):
\begin{equation}
  B_{W.u.}(M1;J_i \rightarrow{}J_f)
  = \frac{B(M1;J_i \rightarrow{}J_f)}{B(W;M1)} \,,
  \label{eq:Wu}
\end{equation}
where $B(W;M1)=(45/8\pi)\mu_N^2$ is the reduced probability of the
nuclear $M$1 transition in the Weisskopf model and $\mu_N$ is the
nuclear magneton.

Nonetheless, the $^{233}$U and $^{229}$Th nuclei are different and
those differences could be crucial, which becomes evident when
comparing to other nuclei with similar level structure. For
example, a $3/2^+3/2[631]\rightarrow 5/2^+5/2[633]$ transition
with an energy of 221.4 keV also exists in the $^{231}$Th nucleus
\cite{Browne-13}. The half life of the $3/2^+3/2[631](221.4$ keV)
state in $^{231}$Th is less than 74 ps and only one gamma
transition, namely, the transition to the ground state has been
observed experimentally from this level, with an internal
conversion coefficient of 1.96 \cite{Browne-13}. This is not
surprising since according to the level scheme \cite{Browne-13},
the quantum numbers of states lying between the
$3/2^+3/2[631](221.4$ keV) level and the ground state are such
that the intensity of other possible transitions must be orders of
magnitude smaller than the $3/2^+3/2[631](221.4$ keV$)\rightarrow
5/2^+5/2[633](0.0)$ transition. Therefore, the transition directly
to the ground state should give the dominant contribution to the
decay of the level in the $^{231}$Th nucleus, and the other decay
channels cannot significantly change the lifetime of the level.
Using the data from Ref.~\cite{Browne-13} the reduced probability
of this transition in the $^{231}$Th nucleus is estimated as
$B_{W.u.}(M1;3/2^+3/2[631](221.4$ keV$) \rightarrow
5/2^+5/2[633](0.0)) \geq 0.93\times 10^{-2}$. This value is at
least three times larger than would have been expected estimating
it from the similar transition in $^{233}$U. Accordingly,
interpolation from the $^{233}$U nucleus to the $^{229}$Th nucleus
could lead to similar results. In addition, it is not obvious that
measurements of the nuclear lifetimes, $\gamma$-ray intensities,
the probabilities of electronic conversion, and other
characteristics of this transition in the $^{233}$U nucleus are
more accurate than for the $^{229}$Th nucleus, where measurement
errors, as we shall see below, are significant.

\begin{figure}
  \includegraphics[angle=270,width=\hsize,keepaspectratio]{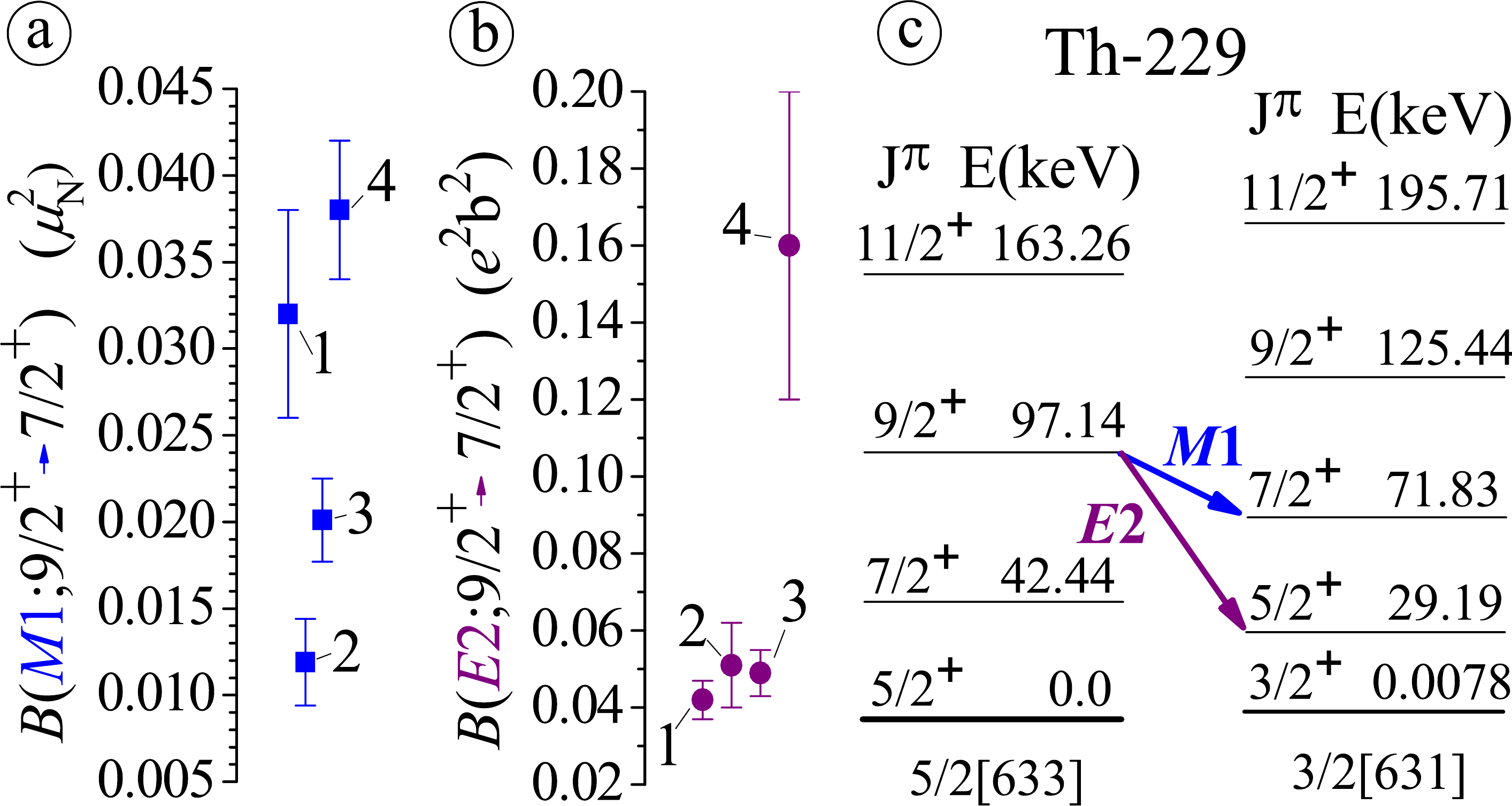}
  \caption{(color online). The experimental values for the reduced probability of the nuclear
    (a) $9/2^+ (97.14$ keV)$ \rightarrow 7/2^+ (71.83$ keV), and (b)
    $9/2^+ (97.14$ keV)$ \rightarrow 5/2^+ (29.19$ keV)  transitions
    The relevant level scheme of the $^{229}$Th nucleus is shown in part (c).
    Data for the transitions were taken from: 1 -- \cite{Bemis-88},
    2 -- \cite{Gulda-02}, 3 -- \cite{Barci-03}, and 4 -- \cite{Ruchowska-06}.}
  \label{fig:M1-E2_Experiment}
\end{figure}

For these reasons, we prefer to use the second approach to
determine an estimate of $B(M1;3/2^+3/2[631](7.8$~eV)$ \rightarrow
5/2^+5/2[633](0.0))$, which relies on available experimental data
for the reduced probability of the $M$1 transitions between the
rotational bands $K^{\pi}[Nn_Z\Lambda]=5/2^+[633]$ and
$3/2^+[631]$ in the $^{229}$Th nucleus and the Alaga rules. Such a
calculation assumes that the adiabatic condition is fulfilled (see
\cite{Dykhne-98_ME} for a detailed analysis of the use of the
adiabatic condition). Here, we do not consider the effects of
nonadiabaticity because of the relatively large uncertainties and
disagreements of the experimental data (see
Fig.~\ref{fig:M1-E2_Experiment}). Further, the well-expressed
rotational structure of the bands in the $^{229}$Th nucleus and a
number of other factors \cite{Dykhne-98_ME} allow us to neglect
the Coriolis interaction for a preliminary estimation of the
reduced probability of the isomeric transition from the
experimental data for the $9/2^+5/2[ 633](97.14$~keV$) \rightarrow
7/2^+3/2[631](71.83$~keV) transition in the $^{229}$Th nucleus. In
this limit, we can, however, provide an estimate of the effect of
the Coriolis interaction, which is quite small, on the matrix
element \cite{Dykhne-98_ME}.

Experimental data for $B(M1;9/2^+5/2[633] (97.14$ keV$)
\rightarrow 7/2^+3/2[631] (71.83$ keV)) are, to our knowledge,
available from four separate experiments \cite{Bemis-88, Gulda-02,
Barci-03, Ruchowska-06}. As can be seen in
Fig.~\ref{fig:M1-E2_Experiment}, the reported values for the $M$1
transition show considerable spread. For comparison, in the case
of the $E$2 transition, there is  consensus between three of the
measurements.

%
%

Using the Alaga rules, it is straightforward to obtain the reduced
probability of the  isomeric nuclear transition in terms of the
measured $M$1 reduced probability:
\begin{multline*}
  B(M1;3/2^+(7.8~{\text{eV}}) \rightarrow 5/2^+(0.0))\\
  = \frac{15}{7} B(M1;9/2^+(97.14~{\text{keV}}) \rightarrow
7/2^+(71.83~{\text{keV}})) \,.
\end{multline*}
The results of this calculation are shown in Table~\ref{tab:B:M1}
for each measured value of $B(M1;9/2^+5/2[633] (97.14$ keV$)
\rightarrow 7/2^+3/2[631] (71.83$ keV)) from Fig.
\ref{fig:M1-E2_Experiment}(a).

\begin{table}
  \caption{Calculated reduced probabilities
    $B_{W.u.}(M1;3/2^+(7.8~{\text{eV}}) \rightarrow 5/2^+(0.0))$ in $^{229}$Th
    based on $B(M1;9/2^+5/2[633] (97.14$ keV$) \rightarrow 7/2^+3/2[631]
    (71.83$ keV) from given references.}
  \label{tab:B:M1}

  \begin{tabular}{c@{\quad}c}
    \hline
    $B_{W.u.}$ ($10^{-2}$) & based on\\
    \hline
    \hline
    $3.83\pm0.72$ & Ref.~\cite{Bemis-88}\\
    $1.42\pm0.30$ & Ref.~\cite{Gulda-02}\\
    $2.41\pm0.29$ & Ref.~\cite{Barci-03}\\
    $4.55\pm0.48$ & Ref.~\cite{Ruchowska-06}\\
    \hline
  \end{tabular}
\end{table}

Interestingly, the data of \cite{Barci-03} affords another means
to obtain $B_{W.u.}(M1;3/2^+(7.8~{\text{eV}}) \rightarrow
5/2^+(0.0))$. In that work, the relative intensities of the
transitions from the level $9/2^+3/2[631](125.44~{\text{keV}})$
were also measured to states $9/2^+(97.14~{\text{keV}})$ and
$7/2^+(42.44~{\text{keV}})$ of the $5/2^+[633]$ rotational band.
This allows us to calculate the reduced probabilities
$B(M1;9/2^+3/2[631](125.44~{\text{keV}}) \rightarrow
9/2^+5/2[633](97.14~{\text{keV}})) = (0.56\pm0.25)\times10^{-2}$
$\mu_N^2$ and $B(M1;9/2^+3/2[631](125.44~{\text{keV}}) \rightarrow
7/2^+5/2[633](42.44~{\text{keV}})) = (0.9\pm0.4)\times10^{-3}$
$\mu_N^2$ in the frame of the rotational model. Using the Alaga
rules, we can then calculate the reduced probability of the
$M$1(7.8~eV) isomeric transition. Both of the reduced
probabilities give practically the same value $B_{W.u.}(M1;
3/2^+(7.8~{\text{eV}}) \rightarrow 5/2^+(0.0)) =
(0.74\pm0.33)\times10^{-2}$. This result is shown in
Table~\ref{tab:B:M1:U:Th} along with the reduced probabilities of
similar transitions in $^{233}$U at 311.9 keV and $^{231}$Th at
221.4 keV.

\begin{table}
  \caption{Reduced probabilities
    $B_{W.u.}(M1;3/2^+3/2[631] \rightarrow 5/2^+5/2[633])$ for other
    nuclei or, for the case of $^{229}$Th, calculated from reduced probabilities
    of other transitions (see text).}
  \label{tab:B:M1:U:Th}
  \begin{tabular}{c@{\quad}c@{\quad}c}
    \hline
    $B_{W.u.}$ ($10^{-2}$) & nucleus & from/based on\\
    \hline
    \hline
    $0.33\pm0.05$ & $^{233}$U  & Ref.~\cite{Singh-05} \\
    $> 0.93$      & $^{231}$Th & Ref.~\cite{Browne-13} \\
    \hline
    $0.74\pm0.33$ & $^{229}$Th & Ref.~\cite{Barci-03} \\
    \hline
  \end{tabular}
\end{table}

Thus, these estimates lead to a significant, more than an order of
magnitude, range in the values for the reduced probability of the
isomeric transition of the $^{229}$Th nucleus. However, if, for
the aforementioned reasons, we restrict the estimate to those
values calculated from the
$9/2^+5/2[633](97.14~{\text{keV}})\rightarrow
7/2^+3/2[631](71.83~{\text{keV}})$ transitions the spread in mean
values is within a factor of 3.

\section{Radiative lifetime of the isomeric level}
\label{sec:Lifetime}

Currently, the generally accepted value for the energy of the
isomeric state $3/2^+3/2[631]$ is $7.8\pm 0.5$~eV \cite{Beck-07,
Beck-R}. Since the energy of the isomeric level exceeds, for
example, the ionization potential of the isolated thorium atom,
6.08~eV, the radiative lifetime of the $^{229}$Th isomeric state
is highly dependent on the chemical environment and electronic
conversion is the dominant decay channel \cite{Strizhov-91}. It is
very difficult to directly observe the transition in such
environments, as both the excitation radiation is absorbed by the
electrons and the energy of any conversion electron is very small
(only a few electron volts), making it difficult to detect.
Similarly, it is difficult to predict the lifetime of the isomeric
state for thorium ions in the Th$^{m+}$, $m < 4$ charge state.
Here, for example, the process of decay via the electronic bridge
\cite{Strizhov-91,Porsev-10_3+,Porsev-10_1+} can dominate, which
cannot be calculated without precise knowledge of the nuclear
transition energy and the wave functions of the valence electrons.

%
%
\begin{figure}[]
  \includegraphics[angle=270,width=0.8\hsize,keepaspectratio]{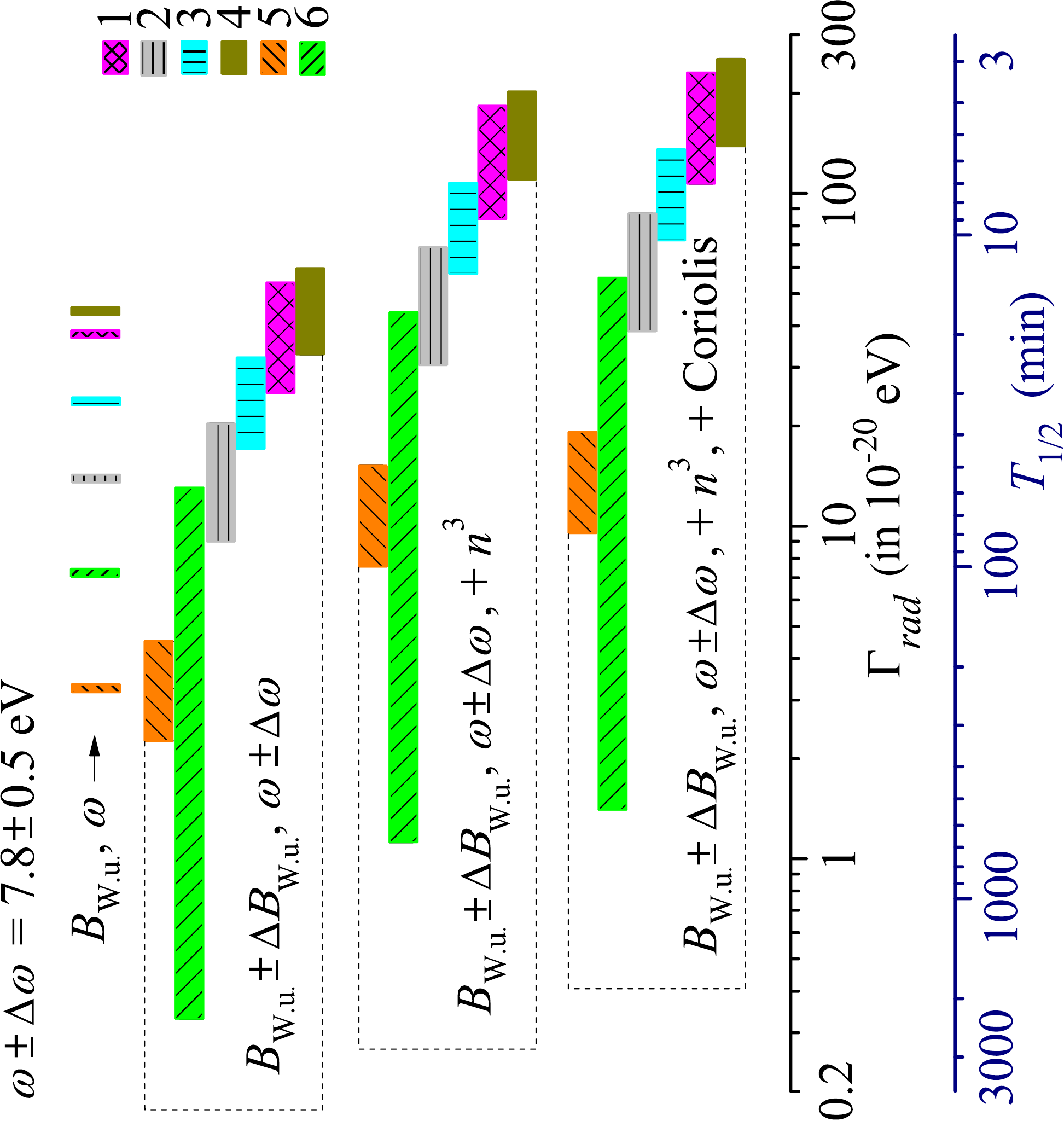}
  \caption{(color online).
    The range of possible radiative linewidths $\Gamma_{rad}$ (upper scale)
    and half-lifes $T_{1/2}$ (lower scale) of the isomeric state
    $3/2^+3/2[631](7.8\pm0.5~{\text{eV}})$ in the $^{229}$Th nucleus.
    Calculations are based on values for
    $B_{W.u.}(M1;3/2^+3/2[631] \rightarrow 5/2^+5/2[633])$ from
    Tab.~\ref{tab:B:M1} (calculated from: 1 -- \cite{Bemis-88},
    2 -- \cite{Gulda-02}, 3 -- \cite{Barci-03}, 4 -- \cite{Ruchowska-06}) and,
    for completeness,
    Tab.~\ref{tab:B:M1:U:Th} (5 -- \cite{Singh-05}; 6 -- \cite{Barci-03}).
    The refractive index $n\approx1.5$  increases the probability of the $M$1
    transition by a factor $n^3$ (third row).
    The Coriolis interaction can lead to a slight increase of the linewidth
    by a factor of 1.2--1.3 \cite{Dykhne-98} (fourth row).}
  \label{fig:T-range}
\end{figure}

In the following, we only estimate the radiative half-life of the
thorium isomeric state in the absence of any chemical effects
based on the reduced probabilities discussed in
Sec.~\ref{sec:MatrixElement}. As discussed previously, we prefer
the four reduced probabilities calculated from the four
measurements for the $M$1 $9/2^+ 5/2[633] (97.14~{\text{keV}})
\rightarrow 7/2^+ 3/2[631] (72.83~{\text{keV}})$ transition in
$^{229}$Th (see Tab.~\ref{tab:B:M1}), which appears most
defensible, as this technique has shown to be accurate to within
experimental error in cases where data exists \cite{Dykhne-98_ME}.
Radiative half-lifes based on the reduced probabilities from other
transitions in $^{229}$Th or the similar transition in $^{233}$U
are only given for completeness (see Tab.~\ref{tab:B:M1:U:Th}).
These results are directly applicable to trapped Th$^{m+}$ ions
with $m\geq4$ and, with a minor modification, to Th in a
large-bandgap crystal. The modification in the latter case is due
to the polarization of the dielectric medium and leads to a
reduction of the half-life by a factor of $1/n^3$
\cite{Rikken-95,Tkalya-00-JETPL}, where $n$ denotes the refractive
index. The calculated half-lifes can further be used in the
trapped ion approach with $m < 4$ to calculate e.g. the electron
bridge process once the electronic spectra of the ions are known.

The results are shown in Fig.~\ref{fig:T-range}, first row, for
$\omega=\unit[7.8]{eV}$. The range of half-lifes including one
standard deviation in both the reduced probabilities and the
currently accepted transition energy is given in the second row.
The calculations for the case of a large-bandgap crystal with a
typical refractive index $n \approx 1.5$ is shown in the third
row. Lastly, the additional inclusion of the Coriolis interaction
leads to only a small correction (fourth row).

Using these results, we construct a favored region for the
radiative half-life as a function of transition energy based only
on the values for the reduced probabilities from
Tab.~\ref{tab:B:M1} (see also Fig.~\ref{fig:Tkalya_Fig5} in
Sec.~\ref{sec:FavoredArea}). The center of this favored region is
defined by the weighted average of the reduced probabilities. The
bounds of the favored region are constructed as 1.96 standard
deviations around the center of the favored region, which
corresponds to roughly a 95\% confidence level, however,
considering that the individual reduced transition probabilities
are not consistent within their errors, we increase the standard
deviation by the Birge ratio of 3.4 \cite{Bodnar-14}:
\begin{equation*}
  0.46\times10^6~{\text{s~eV}}^3/\omega^3 \leq T_{1/2} \leq
  1.5\times10^6~{\text{s~eV}}^3/\omega^3.
\end{equation*}
Here, we did not include a crystal environment, as the inclusion of the
refractive index $n$ is straightforward.  However, the small correction due to
Coriolis interaction is included leading to an increase of the linewidth of the
transition by a factor of 1.2--1.3 \cite{Dykhne-98_ME}.
(These bounds are similar to those of Ref. \cite{Jeet-15}, but we consider
them more accurate as they include e.g. the Birge ratio.)

Alternatively, a more conservative region can be constructed which
is bound by the extreme values of the individual radiative
lifetimes deduced from Tab.~\ref{tab:B:M1}  $\pm1.96$ standard
deviations (see Fig.~\ref{fig:Tkalya_Fig5}).  Its functional form
is given as (again including correction due to Coriolis
interaction, but without crystal environment):
\begin{equation*}
  0.31\times10^6~{\text{s~eV}}^3/\omega^3 \leq T_{1/2} \leq
  2.1\times10^6~{\text{s~eV}}^3/\omega^3.
\end{equation*}

\section{Importance of the electronic conversion decay channel}
\label{sec:Conversion}

As mentioned in Sec.~\ref{sec:Lifetime}, the chemical environment
can significantly affect the half-life of the isomeric state
\cite{Strizhov-91}.
It most likely  explains why, given the currently accepted
value for the isomeric transition energy, that many previous experiments
performed in powders, solids and solutions produced null results
\cite{Browne-01,Kikunaga2009,Mitsugashira2003,Zimmermann2010,Swanberg2012};
similarly, non-VUV
sensitive measurements could have been affected by internal conversion
\cite{Utter1999,Shaw1999}.
The internal conversion process could also have strong implications
for the experiments reported in Refs.~\cite{Zhao-12} and
\cite{Yamaguchi-15}.
Though crystalline material is used as host
in these experiments, the charge state of the thorium atom is not
known, since the thorium atoms are either implanted into
\cite{Zhao-12} or chemically adsorbed onto the surface of
\cite{Yamaguchi-15} the crystal. Therefore, it is likely that
some, if not all, of the thorium atoms are in a local chemical
environment that experiences electronic conversion. As we will see
in the following, it is unlikely the isomeric transition can be
detected in such a system, if internal conversion is present.
Thus, as aptly pointed out in Ref.~\cite{Yamaguchi-15}, any
conclusions drawn from experiments of this type should be
considered preliminary until the thorium chemical environment is
known.

If the energy of the isomeric level $3/2^+3/2[631]$ in the
$^{229}$Th exceeds the binding energy of electrons in the local
chemical environment, the main channel of decay is electronic
conversion \cite{Strizhov-91}. Therefore, the lifetime of the
isomeric state can be significantly reduced compared to the
radiative lifetime only. In the following, we consider electronic
conversion of the isomeric state in the neutral Th atom as an
example to give a rough estimate of the lifetime for the
$^{229}$Th isomeric state in such a chemical environment.

Electronic $M$1 and $E$2 conversion from the valence 6$d$ and 7$s$ shells of
the thorium atom is possible for the nuclear isomeric transition. The
ratio of radiation widths of the $E$2 and $M$1 transitions with energy of 7.8
eV in the $^{229}$Th nucleus is
$\Gamma_{rad}(W;E2)/\Gamma_{rad}(W;M1)\simeq10^{-13}$ in the Weisskopf model,
i.e. when the nuclear reduced probabilities are $B_{W.u.}(M1)=1$  and
$B_{W.u.}(E2)=1$. Therefore, we can neglect the $E$2 contribution to the radiation width
of the level for true values in the ranges $B_{W.u.}(M1;3/2^+(7.8~{\text{eV}})
\rightarrow 5/2^+(0.0))\simeq 10^{-1}$--$10^{-3}$  and
$B_{W.u.}(E2;3/2^+(7.8~{\text{eV}}) \rightarrow 5/2^+(0.0))\simeq 1$--10,
respectively. As for the conversion decay channel, our calculations for the
thorium atom give the relation
$\Gamma_{conv}(W;E2)/\Gamma_{conv}(W;M1)\simeq10^{-6}$ in the Weisskopf model.
Accordingly, for the true range of values for the reduced probabilities we find
$\Gamma_{conv}(E2)/\Gamma_{conv}(M1) \leq 10^{-3}$ for a
transition with energy of 7.8~eV in the $^{229}$Th nucleus. Thus, we
neglect the contribution of electronic $E$2 conversion to the isomeric state
conversion lifetime.

\begin{figure}[]
  \includegraphics[angle=270,width=0.8\hsize,keepaspectratio]{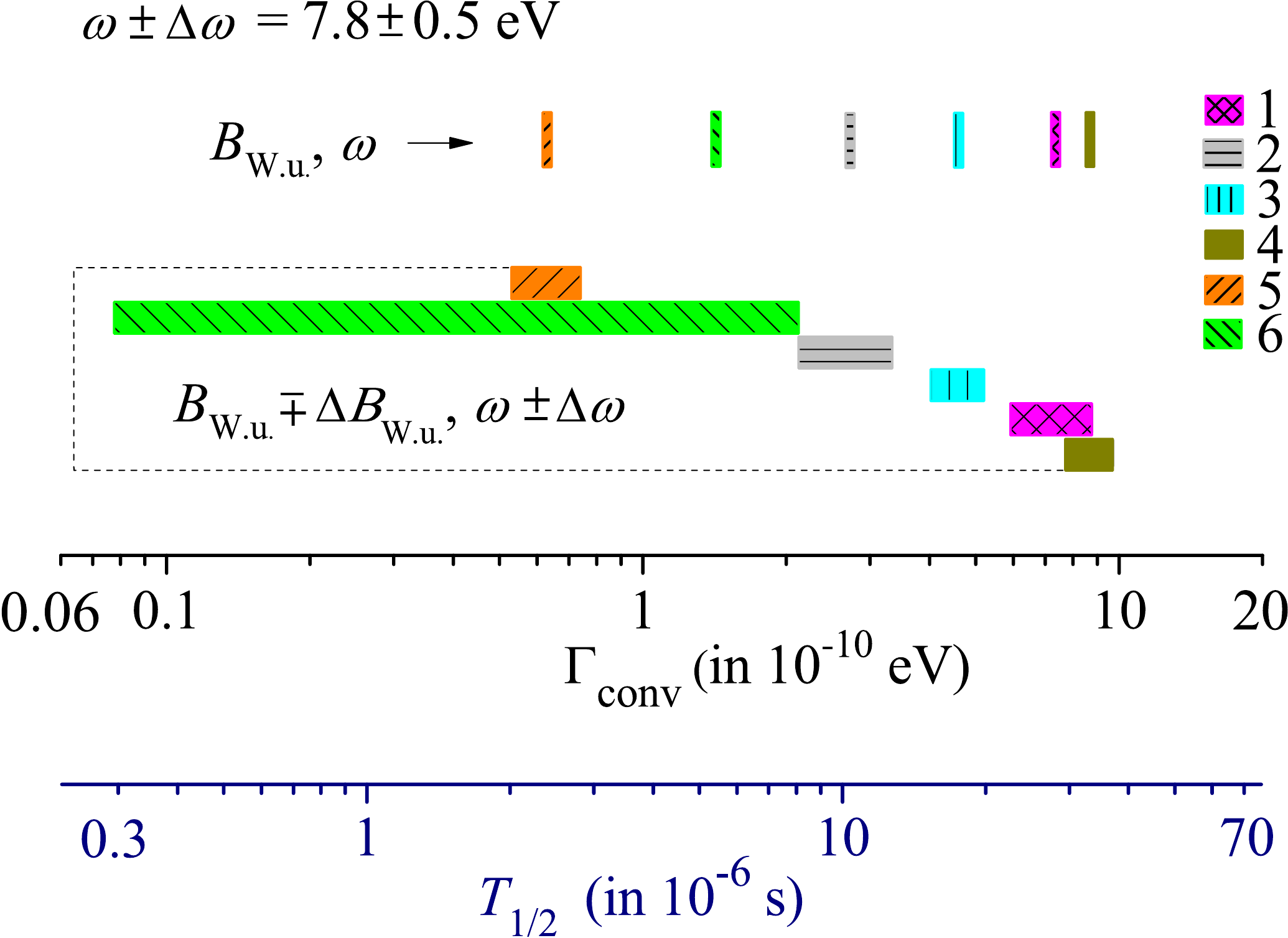}
  \caption{(color online). The ranges of possible conversion widths (the upper scale) and
    lifetimes (the lower scale) of the isomeric nuclear state in a neutral
    isolated Th atom.
    The designations are identical with designations in Fig.~\ref{fig:T-range}.}
  \label{fig:Conversion-range}
\end{figure}

The calculation of the probability $\Gamma_{conv}$ was performed
using code developed in \cite{Soldatov-79} on the basis of known
code in \cite{Band-79}, and then advanced in \cite{Strizhov-91}.
The calculated electronic $M$1 conversion probability for the
energy range 7.3~eV -- 8.3~eV, using the reduced probabilities
from Tables~\ref{tab:B:M1} and \ref{tab:B:M1:U:Th}, are presented
in Figure~\ref{fig:Conversion-range}. Taking into account the
uncertainties on the magnitude of the nuclear matrix element of
the transition, the characteristic lifetime of the isomer in an
atom is $\sim \unit[10]{\mu s}$. As a result, electronic
conversion completely quenches the isomeric state non-radiatively.
Thus, experiments looking for the emitted photons as a signal of
the isomeric transition must ensure that the local chemical
environment of the thorium atom does not support electronic
conversion.

However, this is only one side of the issue. The $M$1 isomeric
transition like the other $M$1 transitions between states of the
bands $K^{\pi}[Nn_Z\Lambda]=5/2^+[633]$ and $3/2^+[631]$, is
first-order forbidden by the asymptotic quantum numbers of the
Nilsson model \cite{Nilsson-55}. Indeed, we are considering $M$1
transitions where $\Delta{}K=1$, $\Delta{}N=0$, $\Delta{}n_Z=0$,
and $\Delta\Lambda=2$, while the selection rules for the $M$1
transition allow the following change for the asymptotic quantum
numbers for the case $\Delta{}K=1$: $\Delta{}N_a=0,\pm2$,
$\Delta{}n_{Z_a}=0,\pm1$, and $\Delta\Lambda_a=0,\pm1$
\cite{Nilsson-55} (the index $a$ means ``allowed''). Thus, the
interband $M$1 transitions $K^{\pi}[Nn_Z\Lambda]=5/2^+[633]
\leftrightarrow 3/2^+[631]$  are weakly forbidden transitions by
the number $\Lambda$. Phenomenology shows that the intensity of
such $M$1 interband transitions are weakened by a factor
$10^{-2n}$,  where $n=|\Delta{}N_a-\Delta{}N| +
|\Delta{}n_{Z_{a}}-\Delta{}n_Z| +
|\Delta{}\Lambda_{a}-\Delta{}\Lambda|$ for $K$-allowed
transitions. In our case $n=1$ and we can expect that such
transitions have reduced probabilities $B_{W.u.}(M1;5/2^+[633]
\leftrightarrow 3/2^+[631])\simeq 10^{-2}$.

The so-called anomalous internal conversion or the dynamic nuclear
volume effect in internal conversion \cite{Church-56} is possible
in transitions forbidden by the asymptotic selection rules. Its
essence is as follows. Amplitudes of the electron wave functions
for the $ms_{1/2}$ $(m=1,2,3...)$ and $lp_{1/2}$ $(l=2,3,4...)$
(or $K$, $L_{I,II}$, $M_{I,II}$...) shells inside the nucleus
differ from zero: $|\psi_{ms_{1/2},lp_{1/2}}(0)|>0$.  In internal
conversion via these shells, the electron current effectively
penetrates into the nucleus, and an ``intranuclear'' internal
conversion becomes possible. The new phenomenon arises if the
coordinates of the electron current $j_e(r)$ and nuclear one
$J_N(R)$ satisfy the condition $r<R<R_0$ (where $R_0$ is a nuclear
radius). In this case, the intranuclear matrix element is changed.
The new nuclear matrix element is not forbidden by the asymptotic
quantum numbers \cite{Voikhansky-66}, and the intranuclear
anomalous conversion becomes possible. Usually, intranuclear
electron conversion is very small and amounts to $(R_0/a_B)^3$,
where $a_B$ is the Bohr radius, as compared with the usual
internal conversion that is gained in the atomic shell outside the
nucleus. But, in the case where the normal nuclear matrix element
is forbidden by the asymptotic quantum numbers of the Nilsson
model and anomalous intranuclear matrix element is allowed by the
asymptotic quantum numbers, the smallness introduced by the
function $(R_0/a_B)^3$ is compensated since the factor $10^{-2n}$
is absent for the anomaly, and anomalous internal conversion
becomes observable.

In this sense, significant difference in the internal conversion
coefficients with the $M$ shell for the $M$1(29.19~keV) transitions
in the $^{229}$Th nucleus, if they really exists, may indicate a
strong anomaly. As aforementioned, the interband transition
provides less than 10\% of the total intensity of the 29.19 keV
transitions. If this transition provides the observed difference in
the internal conversion coefficients, the anomaly probably exists.
And in this case, it will manifest itself in the conversion decay
of the isomeric state $3/2^+(7.8$~eV), because the $7s_{1/2}$
shell is involved to the process. The amplitudes of course obey
the condition $\psi_{7s_{1/2}}(0) \ll
\psi_{2s_{1/2},3s_{1/2}}(0)$. However, the factor
$(\lambda_{is}/a_B)^2$, where $\lambda_{is}= 2\pi/E_{is}$, is
included in the formula for the probability of the dynamic effect
of penetration \cite{Tkalya-94}, compensates the smallness of the
amplitude of the $7s_{1/2}$ wave function inside the nucleus.
Thus, if the dynamic effect of penetration really exists in the
$M$1 interband   transitions, it can also have an impact on the
range of the lifetimes of the $3/2^+(7.8$~eV) isomer in the
conversion decay.

Currently, we can only speculate of the possibility of anomalous
internal conversion, since the accuracy of the measurements
\cite{Gulda-02} were not sufficient. Therefore, it would be
extremely useful if precise measurements of the internal
conversion coefficients for interband magnetic dipole transitions
between the bands $K^{\pi}[Nn_Z\Lambda] = 5/2^+[633]$ and
$3/2^+[631]$ especially at the $K$ and $L$ atomic shells were
performed.

\section{Energy of the nuclear isomeric level}
\label{sec:BranchingCoefficients}

The isomeric transition energy is equally as important as the
isomeric transition radiative lifetime to current experiments.
There have been several attempts
\cite{Reich-90,Helmer-94,Guimaraes-05} to infer the isomeric
transition energy from indirect measurements of $\gamma$-ray
transitions in the $^{229}$Th nucleus. Though the recommended
value for the isomeric transition energy has changed considerably
over the last 40 years, the consensus in the field is to accept
the value put forward in Ref.~\cite{Beck-R} of $E_{is} = 7.8 \pm
0.5$~eV, which updates a previous measurement by the same group
\cite{Beck-07} of $E_{is} = 7.6 \pm 0.5$~eV. In the following, we
detail the dependence of this value on the assumed branching
ratios of interband $E$2 transitions in the $^{229}$Th nucleus. We
find there is considerable spread in the available experimental
data, which could have significant affect on the interpretation of
the data of Ref.~\cite{Beck-07}.

\begin{figure}
  \includegraphics[angle=270,width=0.8\hsize,keepaspectratio]{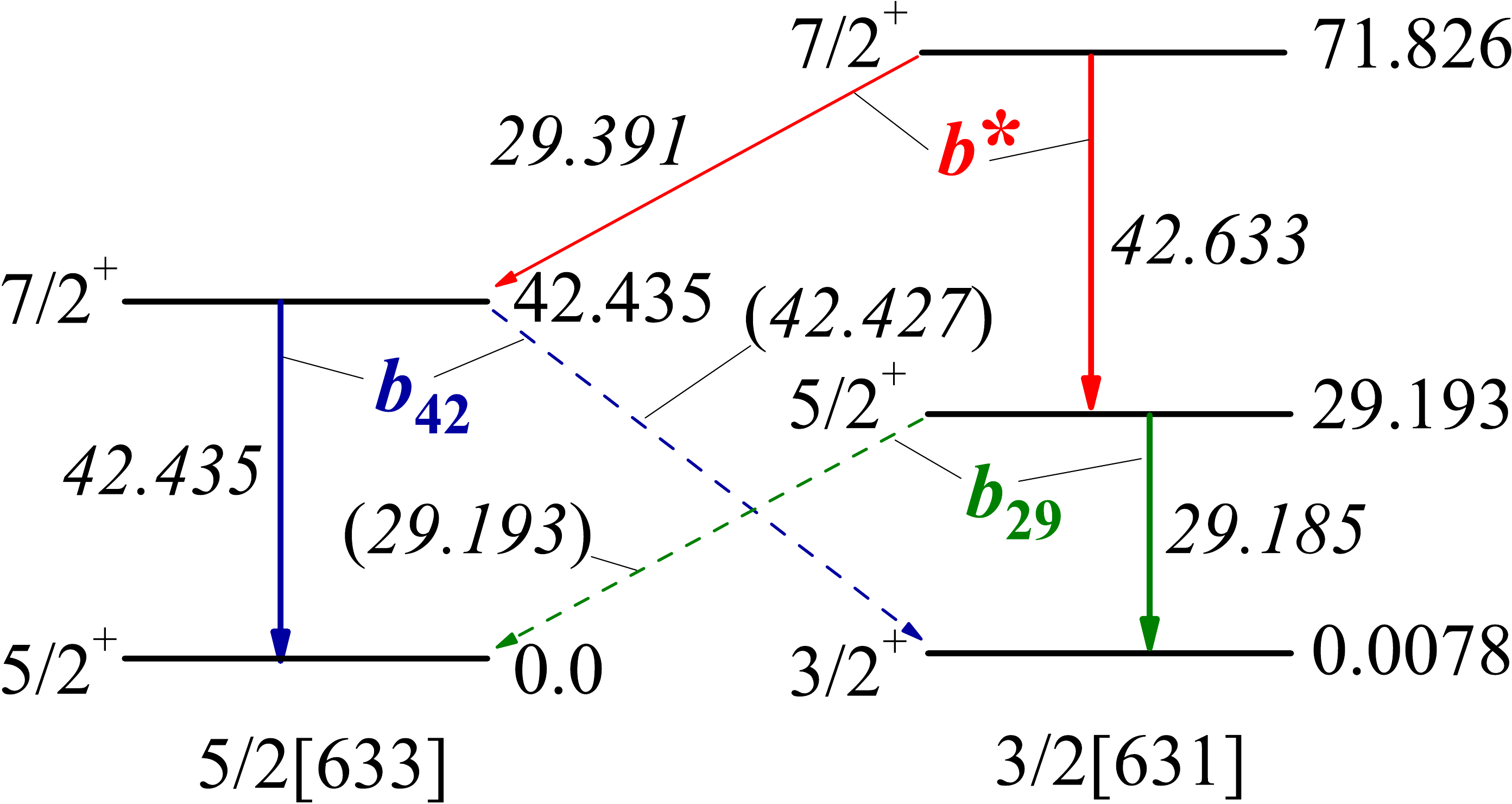}
  \caption{(color online). Relevant part of the level scheme of $^{229}$Th with transitions
    and branching ratios $b^*$, $b_{29}$, and $b_{42}$ used in the works
    \cite{Beck-07,Beck-R} to determine the energy of the isomeric level.
    Energies of the levels and transitions are given in keV.}
  \label{fig:BranchingRatio}
\end{figure}

In their original publication \cite{Beck-07}, $\gamma$-ray
energies of four transitions consisting of a doublet at 29 keV and
a doublet at 42 keV, respectively, were measured with a
state-of-the-art microcalorimeter (see
Fig.~\ref{fig:BranchingRatio}, solid arrows). In their analysis,
they made use of the relation
$E_{is}=\Delta{}E_{29}-\Delta{}E_{42}$, where $\Delta{}E_{29}$ and
$\Delta{}E_{42}$ are the differences between the transition
energies in the corresponding doublets, which reduces the
dependency of the measurement on the absolute calibration of the
detector. The authors accounted for an admixture of the
low-intensity $5/2^+3/2[631](29~{\text{keV}}) \rightarrow
5/2^+5/2[633](0.0)$ transition (see Fig.~\ref{fig:BranchingRatio},
green dashed arrow) to their measured signal of the   transition,
which could not be resolved due to the energy resolution of the
detector (26~eV). This admixture lead to a correction of
$\Delta{}E_{29}-\Delta{}E_{42} = 7.0\pm 0.5$~eV to the value of
the isomeric transition of $E_{is} = 7.6 \pm 0.5$~eV
\cite{Beck-07}. Later, the authors included another unresolved
weak interband transition, $7/2^+5/2[633](42.435~{\text{keV}})
\rightarrow 3/2^+3/2[631](7.8~{\text{eV}})$ (see
Fig.~\ref{fig:BranchingRatio}, blue dashed arrow), in their
analysis, which shifted the isomeric transition energy to the
currently accepted value of $E_{is} = 7.8 \pm 0.5$~eV
\cite{Beck-R}.

%
%

Specifically, the authors showed that the value of the isomeric
transition energy is shifted due to the unresolved transitions by
\cite{Beck-R}
\begin{equation}
  E_{is} = \frac{\Delta{}E_{29}-\Delta{}E_{42}}{1-b_{29}-b_{42}}\,,
  \label{eq:Eis}
\end{equation}
where the branching ratio $b_{29}$ is given as \cite{Beck-07}
\begin{equation}
  b_{29} = \frac{\Gamma_{rad}^{tot}(29.193~{\text{keV}})}%
                {\Gamma_{rad}^{tot}(29.193~{\text{keV}})
         + \Gamma_{rad}^{tot}(29.185~{\text{keV}})}\,,
  \label{eq:b29}
\end{equation}
and the branching ratio $b_{42}$ is given as \cite{Beck-R}
\begin{equation}
  b_{42} = \frac{\Gamma_{rad}^{tot}(42.427~{\text{keV}})}%
                {\Gamma_{rad}^{tot}(42.427~{\text{keV}})
         + \Gamma_{rad}^{tot}(42.435~{\text{keV}})}\,.
  \label{eq:b42}
\end{equation}
In order to determine $b_{29}$, the authors of Ref. \cite{Beck-07}
conducted additional measurements of the branching ratio
\begin{equation}
  b^* = \frac{\Gamma_{rad}^{tot}(29.391~{\text{keV}})}%
             {\Gamma_{rad}^{tot}(29.391~{\text{keV}})
      + \Gamma_{rad}^{tot}(42.633~{\text{keV}})} = \frac{1}{37}
  \label{eq:b*}
\end{equation}
with a quoted measurement accuracy of 8\%. (Here,
$\Gamma_{rad}^{tot} = \Gamma_{rad}(M1) + \Gamma_{rad}(E2)$ and the
designations of the transition energies correspond to those in
Fig.~\ref{fig:BranchingRatio}.) Using the Alaga rules (see
Appendix~\ref{AppendixB}) the authors obtained $b_{29} \approx
1/13$. This led to the aforementioned increase of the energy of
isomeric transition by 0.6~eV in accordance with the
Eq.~(\ref{eq:Eis}). In Ref.~\cite{Beck-R}, the authors performed
an estimation of the value of the $b_{42}$ coefficient, which is
several times smaller (see in Tab.~\ref{tab:b29-b42}) than
$b_{29}$, leading to a smaller shift of 0.2~eV and the currently
accepted value of $E_{is} = 7.8 \pm 0.5$~eV.

Interestingly, the same branching ratios can be obtained from the experimental data~\cite{Bemis-88,Gulda-02,Barci-03,Ruchowska-06}
for parameters of interband (see in Figure~\ref{fig:M1-E2_Experiment}(a) and (b)) and inband
transitions in the rotational bands $5/2^+[633]$ and $3/2^+[631]$
in the $^{229}$Th nucleus. The corresponding
results are given in Table~\ref{tab:b29-b42}, where the
probabilities of the inband transitions were calculated using the
internal quadrupole moment $Q_{20}$ and the difference of the
rotational and internal gyromagnetic ratio $g_R$ and $g_K$,
respectively (for the rotational band $5/2^+[633]$ ---
$Q_{20}=7.1$ $e$b, $|g_K-g_R|=0.176$; for the rotational band
$3/2^+[631]$ --- $Q_{20}=7.1$ $e$b, $|g_K-g_R|=0.56$, (see
\cite{Barci-03})).

\begin{table}
  \caption{Branching ratios $b_{29}$ and $b_{42}$.
           Results are based on the data of the given references.}
  \label{tab:b29-b42}

 \begin{tabular}{c@{\quad}c@{\quad}c@{\quad}c@{\quad}c@{\quad}c@{\quad}c}
    \hline
    Ref.& \cite{Beck-07} & \cite{Beck-R} & \cite{Bemis-88} & \cite{Gulda-02} &
\cite{Barci-03} & \cite{Ruchowska-06}\\
    \hline
    \hline
      $b_{29}$ & 1/13 & & 1/3.5 & 1/7.8 & 1/5.0 & 1/3.1 \\
      $b_{42}$ & & 1/50 & 1/324 & 1/439 & 1/214 & 1/123 \\
    \hline
  \end{tabular}
\end{table}

From Tab.~\ref{tab:b29-b42}, it is obvious that the branching
ratios calculated by the Alaga rules show considerable spread.
This fact is relatively unimportant for the coefficient $b_{42}$,
as the relatively small value given in \cite{Beck-R} is the
largest of the available in Table~\ref{tab:b29-b42}. Thus, if, for
example, $b_{42} = 1/439$ the isomeric transition energy would
effectively shift back to the previous result of $E_{is} = 7.6 \pm
0.5$~eV \cite{Beck-07}. The situation is more dramatic for the
coefficient $b_{29}$. In the scenario $b_{29}\approx 1/3.1$, the
energy of the isomer level would rise up to $\sim10.5$~eV.

The data from Ref.~\cite{Barci-03} allow for two additional
estimates of the coefficient $b_{29}$. Specifically, Table VI of
Ref.~\cite{Barci-03} presents the intensities of gamma transitions
from the levels $7/2^+(71.826$ keV) and $5/2^+(29.193$ keV); some
of these data are experimental results, while others (namely, the
relative intensities of the interband transitions) were calculated
from the strong coupling rotational model. The branching ratios
calculated from these data are presented in Tab.~\ref{tab:b*-b29}.
In the case of the data for decays from the  $7/2^+(71.826$ keV)
level, we calculated the branching ratio $b_{29}$ by the formula
(\ref{eq:B4}) using the coefficient $b^*$.

\begin{table}
  \caption{Coefficients $b^*$ and $b_{29}$ obtained from the data for
  the relative intensities of transitions in \cite{Barci-03}.}
  \label{tab:b*-b29}

 \begin{tabular}{c@{\quad}c@{\quad}c}
    \hline
    Decaying level & $7/2^+(71.826$ keV) & $5/2^+(29.193$ keV) \\
    \hline
    \hline
    $b^*$ & 1/17.5 &  \\
    $b_{29}$ & 1/6.4 & 1/3.9 \\
    \hline
  \end{tabular}
\end{table}

These branching ratios are in rough agreement with those
calculated for Tab.~\ref{tab:b29-b42}, but differ significantly
from the measurement of Ref.~\cite{Beck-07}. The resolution of
this tension between experiments is of the upmost importance. This
can be seen by, for example, taking $b_{29}$ to be given by the
statistical average of the values calculated here. In that case
$b_{29} = 1/5\pm 1/10$ and the isomeric transition energy becomes
$E_{is} = 9.3\pm 1.0$~eV. Despite this disagreement, we cannot
reject the value of $b_{29} = 1/13$ found in Ref.~\cite{Beck-07},
since $b^*$ is given directly from their experimental data and
there is no proof this experiment, which used a state-of-the-art
micro-calorimeter, is less reliable than the other measurements.
Finally, we have performed Monte Carlo simulations of the Beck et
al. experiments and if $b_{29} = 1/3$, an asymmetry in the 29 keV
peak should be visible. Unfortunately, the size of this asymmetry
is smaller than be confirmed by simply viewing the presented date
in Ref.~\cite{Beck-07}. In this regard, it would be interesting to
reanalyze the experimental data.

\section{Defining a favored area and current exclusions}
\label{sec:FavoredArea}

%
%
\begin{figure*}[t]
  \centering
  \begingroup
  \footnotesize
  \setlength{\unitlength}{0.04500bp}%
  \begin{picture}(9600.00,5040.00)%
    \put(0,0){\includegraphics[scale=0.9]{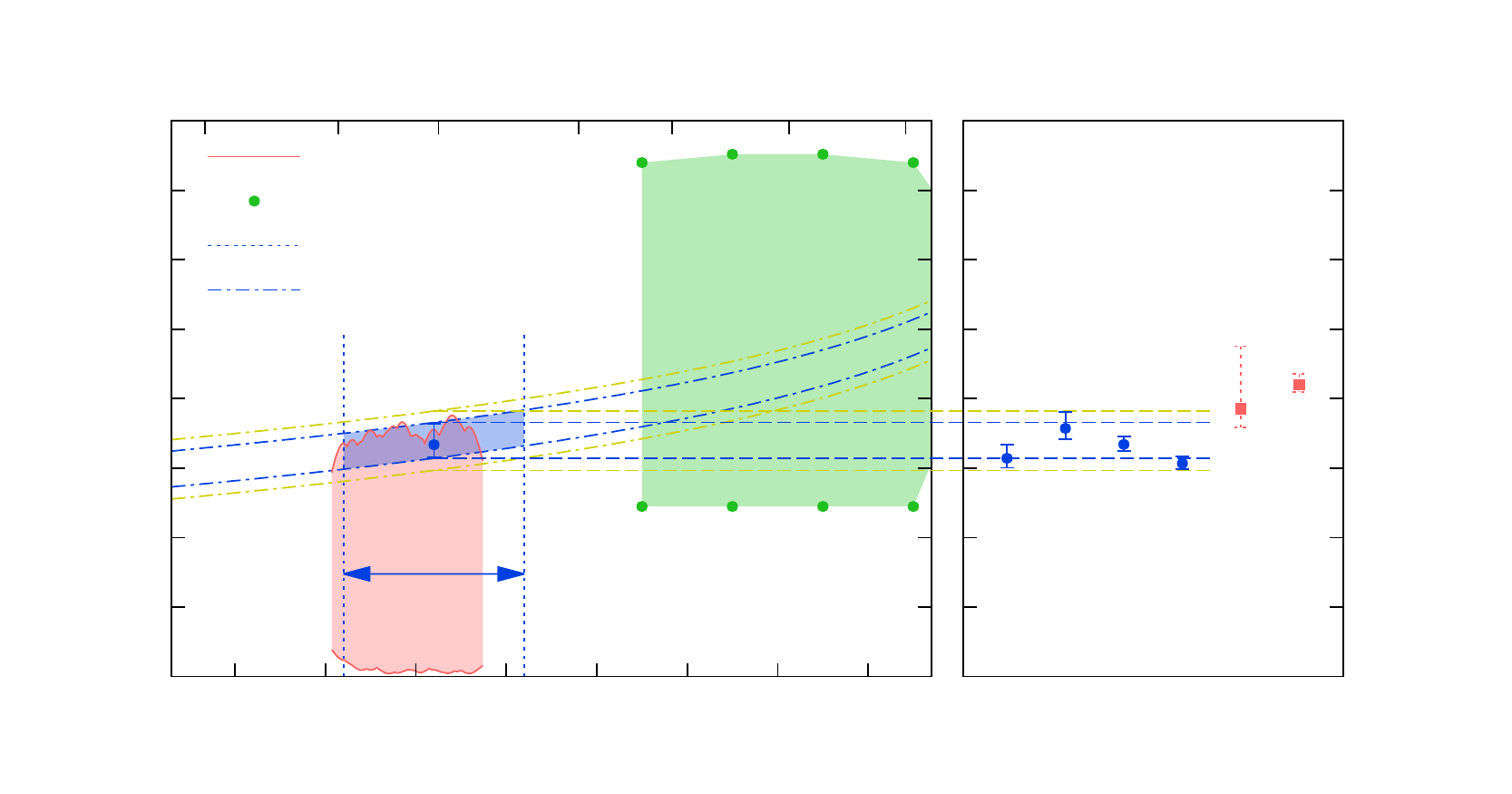}}%
    \put(3378,1407){\makebox(0,0)[l]{\strut{}$\unit[7.8]{eV} \pm 2 \sigma$}}%
    \put(4067,2828){\makebox(0,0){\strut{}$\propto \omega^{-3}$}}%
    \put(396,2519){\rotatebox{-270}{\makebox(0,0){\strut{}lifetime $\tau$ ($\unit{s}$)}}}%
    \put(2014,4056){\makebox(0,0)[l]{\strut{}Ref.~\cite{Jeet-15}}}%
    \put(2014,3774){\makebox(0,0)[l]{\strut{}Ref.~\cite{Moore-04}}}%
    \put(2014,3492){\makebox(0,0)[l]{\strut{}Ref.~\cite{Beck-09}}}%
    \put(2014,3210){\makebox(0,0)[l]{\strut{}favored region}}%
    \put(965,756){\makebox(0,0)[r]{\strut{}$10^0$}}%
    \put(965,1197){\makebox(0,0)[r]{\strut{}$10^1$}}%
    \put(965,1638){\makebox(0,0)[r]{\strut{}$10^2$}}%
    \put(965,2079){\makebox(0,0)[r]{\strut{}$10^3$}}%
    \put(965,2520){\makebox(0,0)[r]{\strut{}$10^4$}}%
    \put(965,2960){\makebox(0,0)[r]{\strut{}$10^5$}}%
    \put(965,3401){\makebox(0,0)[r]{\strut{}$10^6$}}%
    \put(965,3842){\makebox(0,0)[r]{\strut{}$10^7$}}%
    \put(965,4283){\makebox(0,0)[r]{\strut{}$10^8$}}%
    \put(5503,539){\makebox(0,0){\strut{}$3$}}%
    \put(4929,539){\makebox(0,0){\strut{}$4$}}%
    \put(4354,539){\makebox(0,0){\strut{}$5$}}%
    \put(3780,539){\makebox(0,0){\strut{}$6$}}%
    \put(3205,539){\makebox(0,0){\strut{}$7$}}%
    \put(2631,539){\makebox(0,0){\strut{}$8$}}%
    \put(2056,539){\makebox(0,0){\strut{}$9$}}%
    \put(1482,539){\makebox(0,0){\strut{}$10$}}%
    \put(5742,4500){\makebox(0,0){\strut{}$480$}}%
    \put(5001,4500){\makebox(0,0){\strut{}$320$}}%
    \put(4259,4500){\makebox(0,0){\strut{}$240$}}%
    \put(3665,4500){\makebox(0,0){\strut{}$200$}}%
    \put(2775,4500){\makebox(0,0){\strut{}$160$}}%
    \put(2139,4500){\makebox(0,0){\strut{}$140$}}%
    \put(1291,4500){\makebox(0,0){\strut{}$120$}}%
    \put(6942,1407){\makebox(0,0){\strut{}\shortstack{calculated based\\on given Refs.;\\$\unit[95]{\%}$ CL}}}%
    \put(9316,2519){\rotatebox{-270}{\makebox(0,0){\strut{}linewidth $\Gamma$ ($\unit{Hz}$)}}}%
    \put(7313,540){\makebox(0,0){\strut{}$7.8$}}%
    \put(7313,4499){\makebox(0,0){\strut{}$159$}}%
    \put(8634,756){\makebox(0,0)[l]{\strut{}$10^0$}}%
    \put(8634,1197){\makebox(0,0)[l]{\strut{}$10^{-1}$}}%
    \put(8634,1638){\makebox(0,0)[l]{\strut{}$10^{-2}$}}%
    \put(8634,2079){\makebox(0,0)[l]{\strut{}$10^{-3}$}}%
    \put(8634,2520){\makebox(0,0)[l]{\strut{}$10^{-4}$}}%
    \put(8634,2960){\makebox(0,0)[l]{\strut{}$10^{-5}$}}%
    \put(8634,3401){\makebox(0,0)[l]{\strut{}$10^{-6}$}}%
    \put(8634,3842){\makebox(0,0)[l]{\strut{}$10^{-7}$}}%
    \put(8634,4283){\makebox(0,0)[l]{\strut{}$10^{-8}$}}%
    \put(6385,2520){\rotatebox{-270}{\makebox(0,0)[l]{\strut{}~Ref.~\cite{Bemis-88}~}}}%
    \put(6756,2520){\rotatebox{-270}{\makebox(0,0)[l]{\strut{}~Ref.~\cite{Gulda-02}~}}}%
    \put(7127,2520){\rotatebox{-270}{\makebox(0,0)[l]{\strut{}~Ref.~\cite{Barci-03}~}}}%
    \put(7499,2520){\rotatebox{-270}{\makebox(0,0)[l]{\strut{}~Ref.~\cite{Ruchowska-06}~}}}%
    \put(7870,2289){\rotatebox{-270}{\makebox(0,0)[r]{\strut{}~Ref.~\cite{Barci-03}\textsuperscript{*}~}}}%
    \put(8241,2289){\rotatebox{-270}{\makebox(0,0)[r]{\strut{}~Ref.~\cite{Singh-05}~}}}%
    \put(4799,214){\makebox(0,0){\strut{}transition energy $\omega$ ($\unit{eV}$)}}%
    \put(4799,4825){\makebox(0,0){\strut{}wavelength $\lambda$ ($\unit{nm}$)}}%
    \put(1080,171){\makebox(0,0)[l]{\strut{}}}%
  \end{picture}%
  \endgroup
  \caption{(color online). Favored region and experimentally excluded regions for the nuclear
    isomeric transition as a function of transition energy (wavelength) and
    radiative lifetime.
    (Left) The favored region (dash-dotted blue line; see text) and currently
    accepted energy of the isomeric transition (dotted blue lines) according to
    Ref.~\cite{Beck-09} are recommended as primary region of interest for
    current searches (blue shaded area).
    Experimentally excluded areas according to Ref.~\cite{Moore-04} (green
    shaded area between circles) and Ref.~\cite{Jeet-15} (red shaded area
    between solid red lines) exclude parts of the favored region.
    Also shown are the more conservative bounds (outer dash-dotted yellow
    lines; see text).
    The point at 7.8~eV (blue circle) shows exemplarily the weighted average
    of the radiative lifetimes from reduced transition probabilities in
    Tab.~\ref{tab:B:M1} with 1.96 standard deviations including a Birge
    ratio of 3.4.
    (Right)  Radiative lifetimes at 7.8~eV for individual values of the reduced
    transition probability with 1.96 standard deviations according to
    Tab.~\ref{tab:B:M1} (circles, blue) and Tab.~\ref{tab:B:M1:U:Th}
    (squares, red).
    (The latter are only shown for completeness and do not enter into the
    construction of the favored region.)}
  \label{fig:Tkalya_Fig5}
\end{figure*}
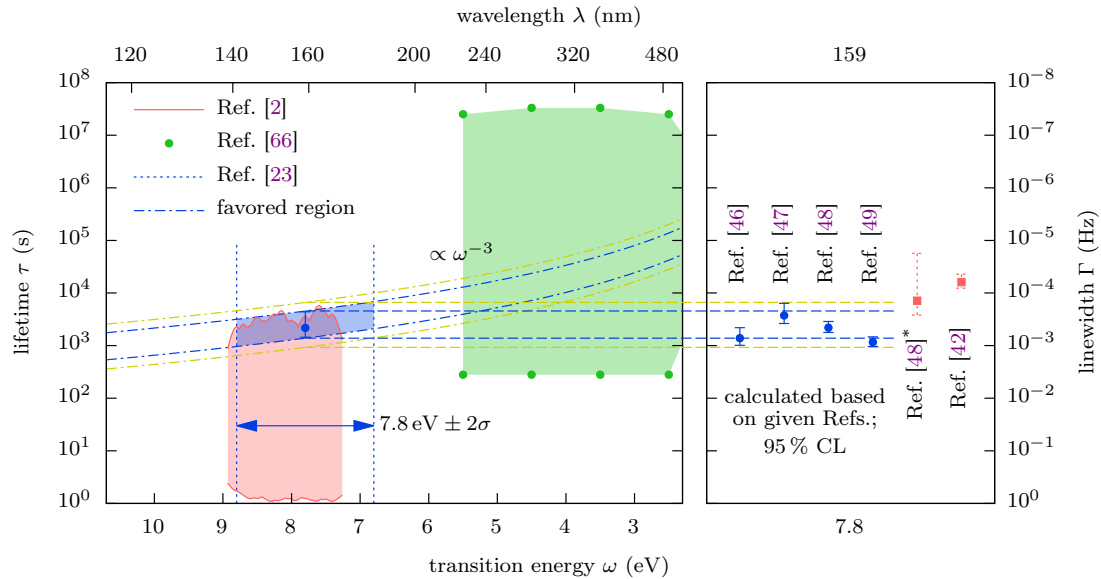

As described in Sec.~\ref{sec:Lifetime}, the measurements of
Refs.~\cite{Bemis-88,Gulda-02,Barci-03,Ruchowska-06}  can be used
to bound the isomeric level half life. In this section we will use
the radiation \emph{lifetime} instead of \emph{half-life}, which
is traditionally used in atomic spectroscopy. The radiative
transition lifetime $\tau$ is bounded by (roughly 95\% confidence
level)
\begin{equation*}
 0.66\times10^6~{\text{s~eV}}^3/\omega^3 \leq\tau\leq
  2.2\times10^6~{\text{s~eV}}^3/\omega^3.
\end{equation*}

In Fig.~\ref{fig:Tkalya_Fig5}, the bound is plotted as a function
of isomeric transition energy (blue dash-dotted lines). For
completeness, the energy ranges from 2.5~eV, which includes the
now-rejected value of the transition energy from
Ref.~\cite{Helmer-94} with one standard deviation towards lower
energies, up to 10.5~eV, which is expected to be the largest
possible value for the transition energy based on
Sec.~\ref{sec:BranchingCoefficients}. Further, the energy range
for the currently accepted value of the transition energy of
Ref.~\cite{Beck-R} including two standard deviations is
highlighted (see Fig.~\ref{fig:Tkalya_Fig5}, blue dotted lines).
The intersection of these two bounds, each at roughly the 95\%
confidence level, is marked as as blue shaded area and gives the
primary region of interest. The energy and lifetime of the
isomeric transition should be found at roughly the 90\% confidence
level in this region.

It is very likely that this region is somewhat conservative in its
upper lifetime bound for two reasons. First, reduced transition
probabilities calculated from the Alaga rules are typically
smaller than actual values when the spin of the nucleus increases
during the transition, as is the case here~\cite{Casten-2000}. Second, the
measurements of the $B(M1;9/2^+5/2[633] (97.14~{\text{keV}})
\rightarrow 7/2^+3/2[631](71.83~{\text{keV}}))$ reduced transition
probability rely on calculated values of the internal conversion
coefficient, and there is evidence \cite{Tretyakov-60} that these
calculated internal conversion coefficients may lead to an
underestimate of the reduced probabilities by a factor of $\sim2$.

Also shown in Fig.~\ref{fig:Tkalya_Fig5}, are the results from
both an indirect \cite{Moore-04} and direct \cite{Jeet-15} search
for the low energy isomeric transition in the $^{229}$Th nucleus;
a similar measurement to \cite{Moore-04} was also performed by
\cite{Kasamatsu2005}.
In Ref.~\cite{Moore-04}, $^{229}$Th produced in the $\alpha$ decay
of $^{233}$U, with an estimated 2\% of the $^{229}$Th populating
the isomeric state, is chemically isolated from the $^{233}$U and
any resulting fluorescence monitored with a photomultiplier tube.
The initial photon count rate, based on a chi-squared analysis of
binned photon counts, is then compared to what is expected from
the known starting amount of $^{233}$U which then sets limits on
the isomer lifetime. From these results, a 99.7\% confidence
interval can be constructed that excludes the possibility of the
transition existing with a certain lifetime in a given energy
range, as depicted by the green shading in
Fig.~\ref{fig:Tkalya_Fig5}.

In the recent direct search \cite{Jeet-15}, tunable, broadband,
synchrotron radiation is used to illuminate a $^{229}$Th$^{4+}$
doped LiSrAlF$_6$ crystal. A photomultiplier tube is used to
detect fluorescence subsequent to illumination of the crystal with
the synchrotron beam. Applying a Feldman-Cousins analysis
\cite{Feldman1998} to the binned photon counts, and comparing this
to what is expected from experimental parameters, an exclusion
region is created, depicted by the red shading between red solid
lines in Fig.~\ref{fig:Tkalya_Fig5}. This exclusion region
represents a 90\% confidence level for the absence of the isomeric
transition.

\section{Results, discussion, and request for future measurements}
\label{sec:Results}

Even a cursory look at the results presented here reveals that the
current situation is far from desirable. There is considerable
scatter in the experimental data leading to a large range of
predictions for the $3/2^+(7.8$~eV) $^{229}$Th isomer radiative
lifetime. Likewise, though the basic value of
$\Delta{}E_{29}-\Delta{}E_{42} = 7.0 \pm 0.5$ ~eV obtained in
\cite{Beck-07} is not currently in question, there is significant
spread in the braching ratio of two unresolved transitions, which
systematically shift the value of the isomeric state.

Of course, the situation could be made much less ambiguous with an
improved measurement of the intensities of the gamma transitions
and internal conversion coefficients between the rotational bands
$5/2^+[633]$ and $3/2^+[631]$ in the $^{229}$Th nucleus. Such an
experiment would ideally resolve the tension between the current
results of Ref.~\cite{Bemis-88,Gulda-02,Barci-03,Ruchowska-06}.
This would allow accurate values of $b_{29}$ and $b_{42}$ to be
calculated. Using these values in Eq.~(\ref{eq:Eis}) and the value
of $\Delta{}E_{29}-\Delta{}E_{42}=7.0\pm 0.5$~eV obtained in
\cite{Beck-07}, which so far is not in question, the isomeric
state energy could be found with greater certainty. Further, these
same experiments would give a definitive means to calculate the
nuclear matrix element of the isomeric transition, and thus the
radiative lifetime of the isomeric state.

Ideally, the measurements should be conducted for transitions
from the rotational band $5/2^+[633]$ to the band $3/2^+[631]$,
and vice versa, in the decay of states of the rotational band
$3/2^+[631]$. Comparison of the data in Tables \ref{tab:B:M1} and
\ref{tab:B:M1:U:Th} indicates a systematic excess of the reduced
probabilities obtained from the analysis of the decay data for the
levels of the $5/2^+[633]$ rotational band. Reduced probabilities
of interband transitions from the rotational band $3/2^+[631]$ to
the band $5/2^+[633]$ are considerably less. This may indicate an
error of measurements, as well as the inapplicability of the
adiabatic approximation in the calculation using the Alaga rules.

If performed, these measurements would considerably sharpen the
region of interest that must be scanned in the search for the
isomeric transition. Given the challenges faced by these searches,
this sharpening of the search space may prove to be a prerequisite
for the completion of this decades old quest.

\section{Acknowledgement}

This work has been supported in part by DARPA (QuASAR program),
ARO (W911NF-11-1-0369), NSF (PHY-1205311), NIST PMG
(60NANB14D302), and RCSA (20112810).

E.T. thanks the Russian government for its support of this work in
the form of the salary of 22 thousand rubles per month (about \$
340) and Moscow State University, which adds to the salary monthly
else 50\% of the specified amount.

\appendix{}
\section{Weisskopf Units}
\label{AppendixA}

In modern nuclear spectroscopy a value of the reduced probability
of transition between the nuclear states with the spins $I_iM_i$
and $I_fM_f$
\begin{equation}
  \begin{split}
    B(E/ML;i\rightarrow{}f) &= \sum_{M_f,M}
      |\langle{}I_fM_f|\hat{\cal{M}}^{E/M}_{LM}|I_iM_i\rangle|^2 \\
    &= \frac{|\langle{}I_f\|\hat{\cal{M}}^{E/M}_{L}\|I_i\rangle|^2}{2I_i+1}\,,
  \end{split}
  \label{eq:A1}
\end{equation}
where $\langle{}I_f\|\hat{\cal{M}}^{E/M}_{L}\|I_i\rangle$ is the
reduced matrix element of the transition operator
$\hat{\cal{M}}^{E/M}_{LM}$, usually is expressed in the Weisskopf
units; see Eq.~(\ref{eq:Wu}) and Ref.~\cite{Blatt-52}. The nuclear
wave functions are as a rule so complex that an exact calculation
of the nuclear matrix elements becomes very difficult. The
simplified single-particle Weisskopf model is convenient because
it makes it easy to evaluate the nuclear matrix element of the
electromagnetic transition. For this purpose the model uses the
proton single-particle radial wave functions of the form $\varphi
= \text{const.}$ inside the nucleus ($R\leq{}R_0$) and $\varphi(R)
= 0$ outside the nucleus ($R>R_0$). The normalization condition
$\int_0^{R_0}|\varphi(R)|^2R^2dR = 1 $ gives $\text{const.} =
\sqrt{3/R_0^3}$. Thus the total wave function of the proton has
the form \cite{Blatt-52}
\begin{equation}
  \Psi({\bf{R}}) = \sqrt{\frac{3}{R_0^3}} Y_{LM}(\Omega_{\bf{R}})\chi_{1/2}
  \quad {\text{for}}\quad{} R\leq{}R_0 \,,
  \label{eq:A2}
\end{equation}
where $\chi_{1/2}$ is the spin part, and $\Psi({\bf{R}}) = 0$ for $R>R_0$.

The radial part of the matrix element of the $EL$ proton transition operator
$\hat{\cal{M}}^{E}_{LM} = eR^LY_{LM}(\Omega)$ is easily calculated with the
wave functions (\ref{eq:A2}):
\begin{equation*}
  \langle\varphi_f(R)|R^L|\varphi_i(R)\rangle = \frac{3}{3+L}R_0^L\,.
  \label{eq:A3}
\end{equation*}

From the angular part of the reduced matrix element in
Eq.~(\ref{eq:A1}), $\langle{}f\|Y_L(\Omega)\|i\rangle$, only a
factor $\sqrt{1/4\pi}$ is left, because
$\sqrt{(2L_i+1)(2L+1)}C^{L_f0}_{L_i0L0}$ with the factor
$1/\sqrt{(2I_i+1)}$ from (\ref{eq:A1}) gives a value of about 1.
(Here $C_{abcd}^{ef}$ is the Clebsch-Gordan coefficient
\cite{Varshalovich-88}.) As a result, the following expression is
obtained in the Weisskopf model for the reduced probability of the
$EL$ single-particle transition
\begin{equation*}
  \begin{split}
    B(W;EL) &= \frac{e^2}{4\pi} \left( \frac{3}{3+L}\right)^2 R_0^{2L} \\
    &= \frac{e^2}{4\pi} \left( \frac{3}{3+L}\right)^2 1.2^{2L}
      A^{2L/3}~{\text{fm}}^{2L}.
  \end{split}
  \label{eq:A4}
\end{equation*}
Here, the value of $R_0=1.2A^{1/3}$ fm was used for the radius of the nucleus
with the atomic number $A$.

For the magnetic $ML$ transition the orbital ($l$) and the spin
($\sigma$) contributions leads to the relation
\begin{equation}
\frac{|\hat{\cal{M}}^{M}_{LM}(l)+\hat{\cal{M}}^{M}_{LM}(\sigma)|^2}{|\hat{\cal{M
}}^{E}_{LM}|^2} \simeq\frac{10}{(M_pR_0)^2} \label{eq:A5}
\end{equation}
between multipole moments of transition \cite{Blatt-52}. ($M_p$ in
Eq.~(\ref{eq:A5}) is the proton mass.) This allows to write for
the reduced probability of magnetic transitions in the Weisskopf
model:
\begin{equation*}
  B(W;ML) = \frac{10}{(M_pR_0)^2}B(W;EL).
  \label{eq:A6}
\end{equation*}

Now, one can express the real reduced probability of a nuclear transition,
$B(M/EL;i\rightarrow{}f)$, through the single particle reduced probability of
the Weisskopf model $B(W;M/EL)$ according
\begin{equation*}
  B(E/ML;i\rightarrow{}f) = B(W;E/ML)B_{W.u.}(E/ML;i\rightarrow{}f) \,,
  \label{eq:A7}
\end{equation*}
where $B_{W.u.}$ is the reduced probability in Weisskopf units, which is
commonly used in tables of nuclear transitions.

The $\gamma$ emission probability in the Weisskopf model
\cite{Blatt-52}
\begin{equation*}
  \Gamma_{rad}(W;E/ML) = \frac{8\pi}{[(2L+1)!!]^2} \frac{L+1}{L} \omega^{2L+1} B(W;E/ML)
  \label{eq:A8}
\end{equation*}
satisfies the condition
\begin{equation*}
  \frac{\Gamma_{rad}(W;L+1)}{\Gamma_{rad}(W;L)} \sim
  (\omega{}R_0)^2 \sim \left(\frac{R_0}{\lambda}\right)^2
  \label{eq:A9}
\end{equation*}
for the emission of $E(L+1)$ and $EL$ (or $M(L+1)$ and $ML$)
multipoles. This approximation is true if the nuclear radius $R_0$
is small compared to the wavelength of nuclear transition
$\lambda$. The latter condition is certainly satisfied for nuclear
transitions with the energies up to several MeV.

\section{Branching Ratios}
\label{AppendixB}

The section discussed the relation of the coefficients $b^*$ and
$b_{29}$. Simple algebraic transformation of the Eq.~(\ref{eq:b*})
enables us to express the ratio of the widths of the transitions
with energies of 42.627 keV and 29.391 keV through coefficient
$b^*$:
\begin{equation}
  \frac{\Gamma_{rad}^{tot}(42.627~{\text{keV}})}%
       {\Gamma_{rad}^{tot}(29.391~{\text{keV}})}
  = \frac{1-b^*}{b^*}
  \label{eq:B1}
\end{equation}
Using the rotational model, we can express the radiation widths of
the 29.185 keV transitions through the widths of the 42.627 keV
transitions:
\begin{align*}
  &\Gamma_{rad}(M1;29.185~{\text{keV}}) \\
    &\:= \left( \frac{C_{5/2\,3/2\,1\,0}^{3/2\,3/2}}%
                  {C_{7/2\,3/2\,1\,0}^{5/2\,3/2}} \right)^2
       \left( \frac{25.185}{42.627} \right)^3
       \times \Gamma_{rad}(M1;42.627~{\text{keV}}) \\
    &\:= 0.240\,\Gamma_{rad}(M1;42.627~{\text{keV}})
\intertext{and}
  &\Gamma_{rad}(E2;29.185~{\text{keV}}) \\
    &\:= \left( \frac{C_{5/2\,3/2\,2\,0}^{3/2\,3/2}}%
                  {C_{7/2\,3/2\,2\,0}^{5/2\,3/2}} \right)^2
       \left( \frac{25.185}{42.627} \right)^5
       \times \Gamma_{rad}(E2;42.627~{\text{keV}}) \\
    &\:= 0.241\,\Gamma_{rad}(E2;42.627~{\text{keV}}).
\end{align*}
Due to the coincidental equality of the transformation
coefficients for the $M$1 and $E$2 components we obtain
\begin{equation}
  \Gamma_{rad}^{tot}(29.185~{\text{keV}})
  = 0.24\,\Gamma_{rad}^{tot}(42.627~{\text{keV}}).
  \label{eq:B2}
\end{equation}

Using the rotational model once again, we obtain the following relations:
\begin{gather*}
  \Gamma_{rad}(M1;29.193~{\text{keV}})
  = 0.735\,\Gamma_{rad}(M1;29.391~{\text{keV}})
\intertext{and}
  \Gamma_{rad}(E2;29.193~{\text{keV}})
  = 1.36\,\Gamma_{rad}(E2;29.391~{\text{keV}}),
\end{gather*}
or, for the total linewidth,
\begin{equation*}
  \begin{split}
    \Gamma_{rad}^{tot}(29.193~{\text{keV}}) &=
    0.735\,\left( \Gamma_{rad}^{tot}(29.391~{\text{keV}}) \right. \\
    &\quad + \left. 0.85\,\Gamma_{rad}(E2;29.391~{\text{keV}}) \right).
  \end{split}
\end{equation*}

Let us estimate the additional part
$0.85\,\Gamma_{rad}(E2;29.391~{\text{keV}})$ to the width
$\Gamma_{rad}^{tot}(29.391~{\text{keV}})$ using the mean values
$B(M1)$ and $B(E2)$ for the interband transitions in
Fig.~\ref{fig:M1-E2_Experiment} (a)--(b). The result is
\begin{equation*}
  \Gamma_{rad}(E2;29.391~{\text{keV}})
  \simeq 2\times10^{-3}\,\Gamma_{rad}(M1;29.391~{\text{keV}}),
\end{equation*}
i.e. we really can neglect by the ``extra'' part
$0.85\,\Gamma_{rad}(E2;29.391~{\text{keV}})$ in the width
$\Gamma_{rad}^{tot}(29.391~{\text{keV}})$ and use the estimation
\begin{equation}
  \Gamma_{rad}^{tot}(29.193~{\text{keV}})
  \approx 0.735\,\Gamma_{rad}^{tot}(29.391~{\text{keV}}).
  \label{eq:B3}
\end{equation}
Substituting expression (\ref{eq:B2}), (\ref{eq:B3}) and
(\ref{eq:B1}) in (\ref{eq:b29}) we finally obtain
\begin{equation}
  \begin{split}
    b_{29} &= \left(1+ \frac{0.24\,\Gamma_{rad}^{tot}(42.627~{\text{keV}})}%
                            {0.735\,\Gamma_{rad}^{tot}(29.193~{\text{keV}})}
              \right)^{-1} \\
           &= \left(1+ \frac{0.24}{0.735} \frac{1-b^*}{b^*} \right)^{-1}.
  \end{split}
  \label{eq:B4}
\end{equation}

\bibstyle{apsrev}
\bibliography{th229_lifetime}

\begin{thebibliography}{71}
\expandafter\ifx\csname natexlab\endcsname\relax\def\natexlab#1{#1}\fi
\expandafter\ifx\csname bibnamefont\endcsname\relax
  \def\bibnamefont#1{#1}\fi
\expandafter\ifx\csname bibfnamefont\endcsname\relax
  \def\bibfnamefont#1{#1}\fi
\expandafter\ifx\csname citenamefont\endcsname\relax
  \def\citenamefont#1{#1}\fi
\expandafter\ifx\csname url\endcsname\relax
  \def\url#1{\texttt{#1}}\fi
\expandafter\ifx\csname urlprefix\endcsname\relax\def\urlprefix{URL }\fi
\providecommand{\bibinfo}[2]{#2}
\providecommand{\eprint}[2][]{\url{#2}}

\bibitem[{\citenamefont{Browne and Tuli}(2008)}]{Browne-08}
\bibinfo{author}{\bibfnamefont{E.}~\bibnamefont{Browne}} \bibnamefont{and}
  \bibinfo{author}{\bibfnamefont{J.~K.} \bibnamefont{Tuli}},
  \bibinfo{journal}{Nucl. Data Sheet} \textbf{\bibinfo{volume}{109}},
  \bibinfo{pages}{2657} (\bibinfo{year}{2008}).

\bibitem[{\citenamefont{Jeet et~al.}(2015)\citenamefont{Jeet, Schneider,
  Sullivan, Rellergert, Mirzadeh, Cassanho, Jenssen, Tkalya, and
  Hudson}}]{Jeet-15}
\bibinfo{author}{\bibfnamefont{J.}~\bibnamefont{Jeet}},
  \bibinfo{author}{\bibfnamefont{C.}~\bibnamefont{Schneider}},
  \bibinfo{author}{\bibfnamefont{S.~T.} \bibnamefont{Sullivan}},
  \bibinfo{author}{\bibfnamefont{W.~G.} \bibnamefont{Rellergert}},
  \bibinfo{author}{\bibfnamefont{S.}~\bibnamefont{Mirzadeh}},
  \bibinfo{author}{\bibfnamefont{A.}~\bibnamefont{Cassanho}},
  \bibinfo{author}{\bibfnamefont{H.~P.} \bibnamefont{Jenssen}},
  \bibinfo{author}{\bibfnamefont{E.~V.} \bibnamefont{Tkalya}},
  \bibnamefont{and} \bibinfo{author}{\bibfnamefont{E.~R.}
  \bibnamefont{Hudson}}, \bibinfo{journal}{Phys. Rev. Lett.}
  \textbf{\bibinfo{volume}{114}}, \bibinfo{pages}{253001}
  (\bibinfo{year}{2015}).

\bibitem[{\citenamefont{Strizhov and Tkalya}(1991)}]{Strizhov-91}
\bibinfo{author}{\bibfnamefont{V.~F.} \bibnamefont{Strizhov}} \bibnamefont{and}
  \bibinfo{author}{\bibfnamefont{E.~V.} \bibnamefont{Tkalya}},
  \bibinfo{journal}{Sov. Phys. JETP} \textbf{\bibinfo{volume}{72}},
  \bibinfo{pages}{387} (\bibinfo{year}{1991}).

\bibitem[{\citenamefont{Porsev and
  Flambaum}(2010{\natexlab{a}})}]{Porsev-10_3+}
\bibinfo{author}{\bibfnamefont{S.~G.} \bibnamefont{Porsev}} \bibnamefont{and}
  \bibinfo{author}{\bibfnamefont{V.~V.} \bibnamefont{Flambaum}},
  \bibinfo{journal}{Phys. Rev. A} \textbf{\bibinfo{volume}{81}},
  \bibinfo{pages}{032504} (\bibinfo{year}{2010}{\natexlab{a}}).

\bibitem[{\citenamefont{Porsev and
  Flambaum}(2010{\natexlab{b}})}]{Porsev-10_1+}
\bibinfo{author}{\bibfnamefont{S.~G.} \bibnamefont{Porsev}} \bibnamefont{and}
  \bibinfo{author}{\bibfnamefont{V.~V.} \bibnamefont{Flambaum}},
  \bibinfo{journal}{Phys. Rev. A} \textbf{\bibinfo{volume}{81}},
  \bibinfo{pages}{042516} (\bibinfo{year}{2010}{\natexlab{b}}).

\bibitem[{\citenamefont{Dicke}(1954)}]{Dicke-54}
\bibinfo{author}{\bibfnamefont{R.~H.} \bibnamefont{Dicke}},
  \bibinfo{journal}{Phys. Rev.} \textbf{\bibinfo{volume}{93}},
  \bibinfo{pages}{99} (\bibinfo{year}{1954}).

\bibitem[{\citenamefont{Tkalya}(2011)}]{Tkalya-11}
\bibinfo{author}{\bibfnamefont{E.~V.} \bibnamefont{Tkalya}},
  \bibinfo{journal}{Phys. Rev. Lett.} \textbf{\bibinfo{volume}{106}},
  \bibinfo{pages}{162501} (\bibinfo{year}{2011}).

\bibitem[{\citenamefont{Flambaum}(2006)}]{Flambaum-06}
\bibinfo{author}{\bibfnamefont{V.~V.} \bibnamefont{Flambaum}},
  \bibinfo{journal}{Phys. Rev. Lett.} \textbf{\bibinfo{volume}{97}},
  \bibinfo{pages}{092502} (\bibinfo{year}{2006}).

\bibitem[{\citenamefont{He and Ren}(2007)}]{He-07}
\bibinfo{author}{\bibfnamefont{H.-t.} \bibnamefont{He}} \bibnamefont{and}
  \bibinfo{author}{\bibfnamefont{Z.-z.} \bibnamefont{Ren}},
  \bibinfo{journal}{J. Phys. G: Nucl. Phys.} \textbf{\bibinfo{volume}{34}},
  \bibinfo{pages}{1611} (\bibinfo{year}{2007}).

\bibitem[{\citenamefont{Hayes and Friar}(2007)}]{Hayes-07}
\bibinfo{author}{\bibfnamefont{A.~C.} \bibnamefont{Hayes}} \bibnamefont{and}
  \bibinfo{author}{\bibfnamefont{J.~L.} \bibnamefont{Friar}},
  \bibinfo{journal}{Phys. Lett. B} \textbf{\bibinfo{volume}{650}},
  \bibinfo{pages}{229} (\bibinfo{year}{2007}).

\bibitem[{\citenamefont{Litvinova et~al.}(2009)\citenamefont{Litvinova,
  Feldmeier, Dobaczewski, and Flambaum}}]{Litvinova2009}
\bibinfo{author}{\bibfnamefont{E.}~\bibnamefont{Litvinova}},
  \bibinfo{author}{\bibfnamefont{H.}~\bibnamefont{Feldmeier}},
  \bibinfo{author}{\bibfnamefont{J.}~\bibnamefont{Dobaczewski}},
  \bibnamefont{and} \bibinfo{author}{\bibfnamefont{V.}~\bibnamefont{Flambaum}},
  \bibinfo{journal}{Phys. Rev. C} \textbf{\bibinfo{volume}{79}},
  \bibinfo{pages}{064303} (\bibinfo{year}{2009}),
  \urlprefix\url{http://link.aps.org/doi/10.1103/PhysRevC.79.064303}.

\bibitem[{\citenamefont{Dykhne and Tkalya}(1998{\natexlab{a}})}]{Dykhne-98}
\bibinfo{author}{\bibfnamefont{A.~M.} \bibnamefont{Dykhne}} \bibnamefont{and}
  \bibinfo{author}{\bibfnamefont{E.~V.} \bibnamefont{Tkalya}},
  \bibinfo{journal}{JETP Lett.} \textbf{\bibinfo{volume}{67}},
  \bibinfo{pages}{549} (\bibinfo{year}{1998}{\natexlab{a}}).

\bibitem[{\citenamefont{Dykhne et~al.}(1996)\citenamefont{Dykhne, Eremin, and
  Tkalya}}]{Dykhne-96}
\bibinfo{author}{\bibfnamefont{A.~M.} \bibnamefont{Dykhne}},
  \bibinfo{author}{\bibfnamefont{N.~V.} \bibnamefont{Eremin}},
  \bibnamefont{and} \bibinfo{author}{\bibfnamefont{E.~V.}
  \bibnamefont{Tkalya}}, \bibinfo{journal}{JETP Lett.}
  \textbf{\bibinfo{volume}{64}}, \bibinfo{pages}{345} (\bibinfo{year}{1996}).

\bibitem[{\citenamefont{Tkalya et~al.}(1996)\citenamefont{Tkalya, Varlamov,
  Lomonosov, and Nikulin}}]{Tkalya-96}
\bibinfo{author}{\bibfnamefont{E.~V.} \bibnamefont{Tkalya}},
  \bibinfo{author}{\bibfnamefont{V.~O.} \bibnamefont{Varlamov}},
  \bibinfo{author}{\bibfnamefont{V.~V.} \bibnamefont{Lomonosov}},
  \bibnamefont{and} \bibinfo{author}{\bibfnamefont{S.~A.}
  \bibnamefont{Nikulin}}, \bibinfo{journal}{Phys. Scr.}
  \textbf{\bibinfo{volume}{53}}, \bibinfo{pages}{296} (\bibinfo{year}{1996}).

\bibitem[{\citenamefont{Peik and Tamm}(2000)}]{Peik-03}
\bibinfo{author}{\bibfnamefont{E.}~\bibnamefont{Peik}} \bibnamefont{and}
  \bibinfo{author}{\bibfnamefont{C.}~\bibnamefont{Tamm}},
  \bibinfo{journal}{Europhys. Lett.} \textbf{\bibinfo{volume}{61}},
  \bibinfo{pages}{181} (\bibinfo{year}{2000}).

\bibitem[{\citenamefont{Rellergert
  et~al.}(2010{\natexlab{a}})\citenamefont{Rellergert, DeMille, Greco, Hehlen,
  Torgerson, and Hudson}}]{Rellergert-10}
\bibinfo{author}{\bibfnamefont{W.~G.} \bibnamefont{Rellergert}},
  \bibinfo{author}{\bibfnamefont{D.}~\bibnamefont{DeMille}},
  \bibinfo{author}{\bibfnamefont{R.~R.} \bibnamefont{Greco}},
  \bibinfo{author}{\bibfnamefont{M.~P.} \bibnamefont{Hehlen}},
  \bibinfo{author}{\bibfnamefont{J.~R.} \bibnamefont{Torgerson}},
  \bibnamefont{and} \bibinfo{author}{\bibfnamefont{E.~R.}
  \bibnamefont{Hudson}}, \bibinfo{journal}{Phys. Rev. Lett.}
  \textbf{\bibinfo{volume}{104}}, \bibinfo{pages}{200802}
  (\bibinfo{year}{2010}{\natexlab{a}}).

\bibitem[{\citenamefont{Campbell et~al.}(2012)\citenamefont{Campbell, Radnaev,
  Kuzmich, Dzuba, Flambaum, and Derevianko}}]{Campbell-12}
\bibinfo{author}{\bibfnamefont{C.~J.} \bibnamefont{Campbell}},
  \bibinfo{author}{\bibfnamefont{A.~G.} \bibnamefont{Radnaev}},
  \bibinfo{author}{\bibfnamefont{A.}~\bibnamefont{Kuzmich}},
  \bibinfo{author}{\bibfnamefont{V.~A.} \bibnamefont{Dzuba}},
  \bibinfo{author}{\bibfnamefont{V.~V.} \bibnamefont{Flambaum}},
  \bibnamefont{and}
  \bibinfo{author}{\bibfnamefont{A.}~\bibnamefont{Derevianko}},
  \bibinfo{journal}{Phys. Rev. Lett.} \textbf{\bibinfo{volume}{108}},
  \bibinfo{pages}{120802} (\bibinfo{year}{2012}).

\bibitem[{\citenamefont{Peik and Okhapkin}(2015)}]{Peik2015}
\bibinfo{author}{\bibfnamefont{E.}~\bibnamefont{Peik}} \bibnamefont{and}
  \bibinfo{author}{\bibfnamefont{M.}~\bibnamefont{Okhapkin}},
  \bibinfo{journal}{C. R. Phys.} \textbf{\bibinfo{volume}{16}},
  \bibinfo{pages}{516} (\bibinfo{year}{2015}), ISSN \bibinfo{issn}{1631-0705},
  \urlprefix\url{http://www.sciencedirect.com/science/article/pii/S1631070515000213}.

\bibitem[{\citenamefont{Burke et~al.}(2008)\citenamefont{Burke, Garrett, Qu,
  and Naumann}}]{Burke-08}
\bibinfo{author}{\bibfnamefont{D.~G.} \bibnamefont{Burke}},
  \bibinfo{author}{\bibfnamefont{P.~E.} \bibnamefont{Garrett}},
  \bibinfo{author}{\bibfnamefont{T.}~\bibnamefont{Qu}}, \bibnamefont{and}
  \bibinfo{author}{\bibfnamefont{R.~A.} \bibnamefont{Naumann}},
  \bibinfo{journal}{Nucl. Phys. A} \textbf{\bibinfo{volume}{809}},
  \bibinfo{pages}{129} (\bibinfo{year}{2008}).

\bibitem[{\citenamefont{Helmer and Reich}(1994)}]{Helmer-94}
\bibinfo{author}{\bibfnamefont{R.~G.} \bibnamefont{Helmer}} \bibnamefont{and}
  \bibinfo{author}{\bibfnamefont{C.~W.} \bibnamefont{Reich}},
  \bibinfo{journal}{Phys. Rev. C} \textbf{\bibinfo{volume}{49}},
  \bibinfo{pages}{1845} (\bibinfo{year}{1994}).

\bibitem[{\citenamefont{Guimaraes-Filho and Helene}(2005)}]{Guimaraes-05}
\bibinfo{author}{\bibfnamefont{Z.~O.} \bibnamefont{Guimaraes-Filho}}
  \bibnamefont{and} \bibinfo{author}{\bibfnamefont{O.}~\bibnamefont{Helene}},
  \bibinfo{journal}{Phys. Rev. C} \textbf{\bibinfo{volume}{71}},
  \bibinfo{pages}{044303} (\bibinfo{year}{2005}).

\bibitem[{\citenamefont{Beck et~al.}(2007)\citenamefont{Beck, Becker,
  Beiersdorfer, Brown, Moody, Wilhelmy, Porter, Kilbourne, and
  Kelley}}]{Beck-07}
\bibinfo{author}{\bibfnamefont{B.~R.} \bibnamefont{Beck}},
  \bibinfo{author}{\bibfnamefont{J.~A.} \bibnamefont{Becker}},
  \bibinfo{author}{\bibfnamefont{P.}~\bibnamefont{Beiersdorfer}},
  \bibinfo{author}{\bibfnamefont{G.~V.} \bibnamefont{Brown}},
  \bibinfo{author}{\bibfnamefont{K.~J.} \bibnamefont{Moody}},
  \bibinfo{author}{\bibfnamefont{J.~B.} \bibnamefont{Wilhelmy}},
  \bibinfo{author}{\bibfnamefont{F.~S.} \bibnamefont{Porter}},
  \bibinfo{author}{\bibfnamefont{C.~A.} \bibnamefont{Kilbourne}},
  \bibnamefont{and} \bibinfo{author}{\bibfnamefont{R.~L.}
  \bibnamefont{Kelley}}, \bibinfo{journal}{Phys. Rev. Lett.}
  \textbf{\bibinfo{volume}{98}}, \bibinfo{pages}{142501}
  (\bibinfo{year}{2007}).

\bibitem[{Bec({\natexlab{a}})}]{Beck-09}
\bibinfo{note}{B. R. Beck, C. Y. Wu, P. Beiersdorfer, G. V. Brown, J. A.
  Becker, J. K. Moody, J. B. Wilhelmy, F. S. Porter, C. A. Kilbourne, and R. L.
  Kelley, Lawrence Livermore National Laboratory, Conference LLNL-PROC-415170,
  2009}, \urlprefix\url{http://www.osti.gov/scitech/biblio/964521-r2Qnkb/#.#}.

\bibitem[{\citenamefont{Zhao et~al.}(2012)\citenamefont{Zhao, de~Escobar,
  Rundberg, Bond, Moody, and Vieira}}]{Zhao-12}
\bibinfo{author}{\bibfnamefont{X.}~\bibnamefont{Zhao}},
  \bibinfo{author}{\bibfnamefont{Y.~N.~M.} \bibnamefont{de~Escobar}},
  \bibinfo{author}{\bibfnamefont{R.}~\bibnamefont{Rundberg}},
  \bibinfo{author}{\bibfnamefont{E.~M.} \bibnamefont{Bond}},
  \bibinfo{author}{\bibfnamefont{A.}~\bibnamefont{Moody}}, \bibnamefont{and}
  \bibinfo{author}{\bibfnamefont{D.~J.} \bibnamefont{Vieira}},
  \bibinfo{journal}{Phys. Rev. Lett.} \textbf{\bibinfo{volume}{109}},
  \bibinfo{pages}{160801} (\bibinfo{year}{2012}).

\bibitem[{\citenamefont{Tkalya}(2000)}]{Tkalya-00-JETPL}
\bibinfo{author}{\bibfnamefont{E.~V.} \bibnamefont{Tkalya}},
  \bibinfo{journal}{JETP Lett.} \textbf{\bibinfo{volume}{71}},
  \bibinfo{pages}{311} (\bibinfo{year}{2000}).

\bibitem[{\citenamefont{Tkalya et~al.}(2000)\citenamefont{Tkalya, Zherikhin,
  and Zhudov}}]{Tkalya-00-PRC}
\bibinfo{author}{\bibfnamefont{E.~V.} \bibnamefont{Tkalya}},
  \bibinfo{author}{\bibfnamefont{A.~N.} \bibnamefont{Zherikhin}},
  \bibnamefont{and} \bibinfo{author}{\bibfnamefont{V.~I.}
  \bibnamefont{Zhudov}}, \bibinfo{journal}{Phys. Rev. C}
  \textbf{\bibinfo{volume}{61}}, \bibinfo{pages}{064308}
  (\bibinfo{year}{2000}).

\bibitem[{\citenamefont{K\"{a}lber et~al.}(1989)\citenamefont{K\"{a}lber, Rink,
  Bekk, Faubel, G\"{o}ring, Meisel, Rebel, and Thompson}}]{Kaelber1989}
\bibinfo{author}{\bibfnamefont{W.}~\bibnamefont{K\"{a}lber}},
  \bibinfo{author}{\bibfnamefont{J.}~\bibnamefont{Rink}},
  \bibinfo{author}{\bibfnamefont{K.}~\bibnamefont{Bekk}},
  \bibinfo{author}{\bibfnamefont{W.}~\bibnamefont{Faubel}},
  \bibinfo{author}{\bibfnamefont{S.}~\bibnamefont{G\"{o}ring}},
  \bibinfo{author}{\bibfnamefont{G.}~\bibnamefont{Meisel}},
  \bibinfo{author}{\bibfnamefont{H.}~\bibnamefont{Rebel}}, \bibnamefont{and}
  \bibinfo{author}{\bibfnamefont{R.}~\bibnamefont{Thompson}},
  \bibinfo{journal}{Z. Phys. A} \textbf{\bibinfo{volume}{334}},
  \bibinfo{pages}{103} (\bibinfo{year}{1989}), ISSN \bibinfo{issn}{0939-7922},
  \urlprefix\url{http://dx.doi.org/10.1007/BF01294392}.

\bibitem[{\citenamefont{Campbell et~al.}(2009)\citenamefont{Campbell, Steele,
  Churchill, DePalatis, Naylor, Matsukevich, Kuzmich, and
  Chapman}}]{Campbell-09}
\bibinfo{author}{\bibfnamefont{C.~J.} \bibnamefont{Campbell}},
  \bibinfo{author}{\bibfnamefont{A.~V.} \bibnamefont{Steele}},
  \bibinfo{author}{\bibfnamefont{L.~R.} \bibnamefont{Churchill}},
  \bibinfo{author}{\bibfnamefont{M.~V.} \bibnamefont{DePalatis}},
  \bibinfo{author}{\bibfnamefont{D.~E.} \bibnamefont{Naylor}},
  \bibinfo{author}{\bibfnamefont{D.~N.} \bibnamefont{Matsukevich}},
  \bibinfo{author}{\bibfnamefont{A.}~\bibnamefont{Kuzmich}}, \bibnamefont{and}
  \bibinfo{author}{\bibfnamefont{M.~S.} \bibnamefont{Chapman}},
  \bibinfo{journal}{Phys. Rev. Lett.} \textbf{\bibinfo{volume}{102}},
  \bibinfo{pages}{233004} (\bibinfo{year}{2009}).

\bibitem[{\citenamefont{Herrera-Sancho
  et~al.}(2012)\citenamefont{Herrera-Sancho, Okhapkin, Zimmermann, Tamm, Peik,
  Taichenachev, Yudin, and Glowacki}}]{Herrera-Sancho-12}
\bibinfo{author}{\bibfnamefont{O.~A.} \bibnamefont{Herrera-Sancho}},
  \bibinfo{author}{\bibfnamefont{M.~V.} \bibnamefont{Okhapkin}},
  \bibinfo{author}{\bibfnamefont{K.}~\bibnamefont{Zimmermann}},
  \bibinfo{author}{\bibfnamefont{C.}~\bibnamefont{Tamm}},
  \bibinfo{author}{\bibfnamefont{E.}~\bibnamefont{Peik}},
  \bibinfo{author}{\bibfnamefont{A.~V.} \bibnamefont{Taichenachev}},
  \bibinfo{author}{\bibfnamefont{V.~I.} \bibnamefont{Yudin}}, \bibnamefont{and}
  \bibinfo{author}{\bibfnamefont{P.}~\bibnamefont{Glowacki}},
  \bibinfo{journal}{Phys. Rev. A} \textbf{\bibinfo{volume}{85}},
  \bibinfo{pages}{033402} (\bibinfo{year}{2012}).

\bibitem[{\citenamefont{Rellergert
  et~al.}(2010{\natexlab{b}})\citenamefont{Rellergert, Sullivan, DeMille,
  Greco, Hehlen, Jackson, Torgerson, and Hudson}}]{Rellergert-10b}
\bibinfo{author}{\bibfnamefont{W.~G.} \bibnamefont{Rellergert}},
  \bibinfo{author}{\bibfnamefont{S.~T.} \bibnamefont{Sullivan}},
  \bibinfo{author}{\bibfnamefont{D.}~\bibnamefont{DeMille}},
  \bibinfo{author}{\bibfnamefont{R.~R.} \bibnamefont{Greco}},
  \bibinfo{author}{\bibfnamefont{M.~P.} \bibnamefont{Hehlen}},
  \bibinfo{author}{\bibfnamefont{R.~A.} \bibnamefont{Jackson}},
  \bibinfo{author}{\bibfnamefont{J.~R.} \bibnamefont{Torgerson}},
  \bibnamefont{and} \bibinfo{author}{\bibfnamefont{E.~R.}
  \bibnamefont{Hudson}}, \bibinfo{journal}{IOP Conf. Ser.: Mater. Sci. Eng.}
  \textbf{\bibinfo{volume}{15}}, \bibinfo{pages}{012005}
  (\bibinfo{year}{2010}{\natexlab{b}}).

\bibitem[{\citenamefont{Hehlen et~al.}(2013)\citenamefont{Hehlen, Greco,
  Rellergert, Sullivan, DeMille, Jackson, Hudson, and Torgerson}}]{Hehlen-13}
\bibinfo{author}{\bibfnamefont{M.~P.} \bibnamefont{Hehlen}},
  \bibinfo{author}{\bibfnamefont{R.~R.} \bibnamefont{Greco}},
  \bibinfo{author}{\bibfnamefont{W.~G.} \bibnamefont{Rellergert}},
  \bibinfo{author}{\bibfnamefont{S.~T.} \bibnamefont{Sullivan}},
  \bibinfo{author}{\bibfnamefont{D.}~\bibnamefont{DeMille}},
  \bibinfo{author}{\bibfnamefont{R.~A.} \bibnamefont{Jackson}},
  \bibinfo{author}{\bibfnamefont{E.~R.} \bibnamefont{Hudson}},
  \bibnamefont{and} \bibinfo{author}{\bibfnamefont{J.~R.}
  \bibnamefont{Torgerson}}, \bibinfo{journal}{J. Lumin.}
  \textbf{\bibinfo{volume}{133}}, \bibinfo{pages}{91} (\bibinfo{year}{2013}).

\bibitem[{\citenamefont{Dessovic et~al.}(2014)\citenamefont{Dessovic, Mohn,
  Jackson, Winkler, Schreitl, Kazakov, and Schumm}}]{Dessovic-14}
\bibinfo{author}{\bibfnamefont{P.}~\bibnamefont{Dessovic}},
  \bibinfo{author}{\bibfnamefont{P.}~\bibnamefont{Mohn}},
  \bibinfo{author}{\bibfnamefont{R.~A.} \bibnamefont{Jackson}},
  \bibinfo{author}{\bibfnamefont{J.}~\bibnamefont{Winkler}},
  \bibinfo{author}{\bibfnamefont{M.}~\bibnamefont{Schreitl}},
  \bibinfo{author}{\bibfnamefont{G.}~\bibnamefont{Kazakov}}, \bibnamefont{and}
  \bibinfo{author}{\bibfnamefont{T.}~\bibnamefont{Schumm}},
  \bibinfo{journal}{J. Phys.: Condens. Matter} \textbf{\bibinfo{volume}{26}},
  \bibinfo{pages}{105402} (\bibinfo{year}{2014}).

\bibitem[{\citenamefont{Stellmer et~al.}(2015)\citenamefont{Stellmer, Schreitl,
  and Schumm}}]{Stellmer2015}
\bibinfo{author}{\bibfnamefont{S.}~\bibnamefont{Stellmer}},
  \bibinfo{author}{\bibfnamefont{M.}~\bibnamefont{Schreitl}}, \bibnamefont{and}
  \bibinfo{author}{\bibfnamefont{T.}~\bibnamefont{Schumm}},
  \bibinfo{journal}{arXiv} \textbf{\bibinfo{volume}{1506.01938}},
  \bibinfo{pages}{1} (\bibinfo{year}{2015}), \eprint{1506.01938},
  \urlprefix\url{http://arxiv.org/abs/1506.01938v1}.

\bibitem[{\citenamefont{Ellis et~al.}(2014)\citenamefont{Ellis, Wen, and
  Martin}}]{Ellis-14}
\bibinfo{author}{\bibfnamefont{J.~K.} \bibnamefont{Ellis}},
  \bibinfo{author}{\bibfnamefont{X.-D.} \bibnamefont{Wen}}, \bibnamefont{and}
  \bibinfo{author}{\bibfnamefont{R.~L.} \bibnamefont{Martin}},
  \bibinfo{journal}{Inorg. Chem.} \textbf{\bibinfo{volume}{53}},
  \bibinfo{pages}{6769} (\bibinfo{year}{2014}).

\bibitem[{\citenamefont{Yamaguchi et~al.}(2015)\citenamefont{Yamaguchi, Kolbe,
  Kaser, Reichel, Gottwald, and Peik}}]{Yamaguchi-15}
\bibinfo{author}{\bibfnamefont{A.}~\bibnamefont{Yamaguchi}},
  \bibinfo{author}{\bibfnamefont{M.}~\bibnamefont{Kolbe}},
  \bibinfo{author}{\bibfnamefont{H.}~\bibnamefont{Kaser}},
  \bibinfo{author}{\bibfnamefont{T.}~\bibnamefont{Reichel}},
  \bibinfo{author}{\bibfnamefont{A.}~\bibnamefont{Gottwald}}, \bibnamefont{and}
  \bibinfo{author}{\bibfnamefont{E.}~\bibnamefont{Peik}}, \bibinfo{journal}{New
  J. Phys.} \textbf{\bibinfo{volume}{17}}, \bibinfo{pages}{053053}
  (\bibinfo{year}{2015}).

\bibitem[{\citenamefont{Okhapkin et~al.}(2015)\citenamefont{Okhapkin, Meier,
  Peik, Safronova, Kozlov, and Porsev}}]{Okhapkin-15}
\bibinfo{author}{\bibfnamefont{M.~V.} \bibnamefont{Okhapkin}},
  \bibinfo{author}{\bibfnamefont{D.~M.} \bibnamefont{Meier}},
  \bibinfo{author}{\bibfnamefont{E.}~\bibnamefont{Peik}},
  \bibinfo{author}{\bibfnamefont{M.~S.} \bibnamefont{Safronova}},
  \bibinfo{author}{\bibfnamefont{M.~G.} \bibnamefont{Kozlov}},
  \bibnamefont{and} \bibinfo{author}{\bibfnamefont{S.~G.}
  \bibnamefont{Porsev}}, \bibinfo{journal}{Phys. Rev. A}
  \textbf{\bibinfo{volume}{92}}, \bibinfo{pages}{020503(R)}
  (\bibinfo{year}{2015}).

\bibitem[{\citenamefont{Campbell et~al.}(2011)\citenamefont{Campbell, Radnaev,
  and Kuzmich}}]{Campbell-11}
\bibinfo{author}{\bibfnamefont{C.~J.} \bibnamefont{Campbell}},
  \bibinfo{author}{\bibfnamefont{A.~G.} \bibnamefont{Radnaev}},
  \bibnamefont{and} \bibinfo{author}{\bibfnamefont{A.}~\bibnamefont{Kuzmich}},
  \bibinfo{journal}{Phys. Rev. Lett.} \textbf{\bibinfo{volume}{106}},
  \bibinfo{pages}{223001} (\bibinfo{year}{2011}).

\bibitem[{\citenamefont{Radnaev et~al.}(2012)\citenamefont{Radnaev, Campbell,
  and Kuzmich}}]{Radnaev-12}
\bibinfo{author}{\bibfnamefont{A.~G.} \bibnamefont{Radnaev}},
  \bibinfo{author}{\bibfnamefont{C.~J.} \bibnamefont{Campbell}},
  \bibnamefont{and} \bibinfo{author}{\bibfnamefont{A.}~\bibnamefont{Kuzmich}},
  \bibinfo{journal}{Phys. Rev. A} \textbf{\bibinfo{volume}{44}},
  \bibinfo{pages}{060501(R)} (\bibinfo{year}{2012}).

\bibitem[{\citenamefont{Beloy}(2014)}]{Beloy-14}
\bibinfo{author}{\bibfnamefont{K.}~\bibnamefont{Beloy}},
  \bibinfo{journal}{Phys. Rev. Lett.} \textbf{\bibinfo{volume}{112}},
  \bibinfo{pages}{062503} (\bibinfo{year}{2014}).

\bibitem[{\citenamefont{Herrera-Sancho
  et~al.}(2013)\citenamefont{Herrera-Sancho, Nemitz, Okhapkin, and
  Peik}}]{HerreraSancho2013}
\bibinfo{author}{\bibfnamefont{O.~A.} \bibnamefont{Herrera-Sancho}},
  \bibinfo{author}{\bibfnamefont{N.}~\bibnamefont{Nemitz}},
  \bibinfo{author}{\bibfnamefont{M.~V.} \bibnamefont{Okhapkin}},
  \bibnamefont{and} \bibinfo{author}{\bibfnamefont{E.}~\bibnamefont{Peik}},
  \bibinfo{journal}{Phys. Rev. A} \textbf{\bibinfo{volume}{88}},
  \bibinfo{pages}{012512} (\bibinfo{year}{2013}),
  \urlprefix\url{http://link.aps.org/doi/10.1103/PhysRevA.88.012512}.

\bibitem[{Bec({\natexlab{b}})}]{Beck-R}
\bibinfo{note}{B. R. Beck and J. A. Becker and P. Beiersdorfer and G. V. Brown
  and K. J. Moody and J. B. Wilhelmy and F. S. Porter and C. A. Kilbourne and
  R. L. Kelley, Report LLNL-PROC-415170.},
  \urlprefix\url{https://e-reports-ext.llnl.gov/pdf/375773.pdf}.

\bibitem[{\citenamefont{Singh and Tuli}(2005)}]{Singh-05}
\bibinfo{author}{\bibfnamefont{B.}~\bibnamefont{Singh}} \bibnamefont{and}
  \bibinfo{author}{\bibfnamefont{J.~K.} \bibnamefont{Tuli}},
  \bibinfo{journal}{Nucl. Data Sheet} \textbf{\bibinfo{volume}{105}},
  \bibinfo{pages}{109} (\bibinfo{year}{2005}).

\bibitem[{Bla()}]{Blatt-52}
\bibinfo{note}{J. M. Blatt and V. F. Weisskopf, {\it{Theoretical Nuclear
  Physics}}. John Wiley \& Sons, Inc. NY, 1952.}

\bibitem[{\citenamefont{Browne and Tuli}(2013)}]{Browne-13}
\bibinfo{author}{\bibfnamefont{E.}~\bibnamefont{Browne}} \bibnamefont{and}
  \bibinfo{author}{\bibfnamefont{J.~K.} \bibnamefont{Tuli}},
  \bibinfo{journal}{Nucl. Data Sheet} \textbf{\bibinfo{volume}{114}},
  \bibinfo{pages}{751} (\bibinfo{year}{2013}).

\bibitem[{\citenamefont{Dykhne and Tkalya}(1998{\natexlab{b}})}]{Dykhne-98_ME}
\bibinfo{author}{\bibfnamefont{A.~M.} \bibnamefont{Dykhne}} \bibnamefont{and}
  \bibinfo{author}{\bibfnamefont{E.~V.} \bibnamefont{Tkalya}},
  \bibinfo{journal}{JETP Lett.} \textbf{\bibinfo{volume}{67}},
  \bibinfo{pages}{251} (\bibinfo{year}{1998}{\natexlab{b}}).

\bibitem[{\citenamefont{C.~E.~Bemis et~al.}(1988)\citenamefont{C.~E.~Bemis,
  McGowan, J.~L. C.~Ford, Milner, Robinson, Stelson, Leander, and
  Reich}}]{Bemis-88}
\bibinfo{author}{\bibfnamefont{J.}~\bibnamefont{C.~E.~Bemis}},
  \bibinfo{author}{\bibfnamefont{F.~K.} \bibnamefont{McGowan}},
  \bibinfo{author}{\bibfnamefont{J.}~\bibnamefont{J.~L. C.~Ford}},
  \bibinfo{author}{\bibfnamefont{W.~T.} \bibnamefont{Milner}},
  \bibinfo{author}{\bibfnamefont{R.~L.} \bibnamefont{Robinson}},
  \bibinfo{author}{\bibfnamefont{P.~H.} \bibnamefont{Stelson}},
  \bibinfo{author}{\bibfnamefont{G.~A.} \bibnamefont{Leander}},
  \bibnamefont{and} \bibinfo{author}{\bibfnamefont{C.~W.} \bibnamefont{Reich}},
  \bibinfo{journal}{Phys. Scr.} \textbf{\bibinfo{volume}{38}},
  \bibinfo{pages}{657} (\bibinfo{year}{1988}).

\bibitem[{\citenamefont{Gulda et~al.}(2002)}]{Gulda-02}
\bibinfo{author}{\bibfnamefont{K.}~\bibnamefont{Gulda}} \bibnamefont{et~al.},
  \bibinfo{journal}{(ISOLDE Collaboration), Nucl. Phys. A}
  \textbf{\bibinfo{volume}{703}}, \bibinfo{pages}{45} (\bibinfo{year}{2002}).

\bibitem[{\citenamefont{Barci et~al.}(2003)\citenamefont{Barci, Ardisson,
  Barci-Funel, Weiss, El~Samad, and Sheline}}]{Barci-03}
\bibinfo{author}{\bibfnamefont{V.}~\bibnamefont{Barci}},
  \bibinfo{author}{\bibfnamefont{G.}~\bibnamefont{Ardisson}},
  \bibinfo{author}{\bibfnamefont{G.}~\bibnamefont{Barci-Funel}},
  \bibinfo{author}{\bibfnamefont{B.}~\bibnamefont{Weiss}},
  \bibinfo{author}{\bibfnamefont{O.}~\bibnamefont{El~Samad}}, \bibnamefont{and}
  \bibinfo{author}{\bibfnamefont{R.~K.} \bibnamefont{Sheline}},
  \bibinfo{journal}{Phys. Rev. C} \textbf{\bibinfo{volume}{68}},
  \bibinfo{pages}{034329} (\bibinfo{year}{2003}).

\bibitem[{\citenamefont{Ruchowska et~al.}(2006)\citenamefont{Ruchowska,
  Plociennik, Zylicz et~al.}}]{Ruchowska-06}
\bibinfo{author}{\bibfnamefont{E.}~\bibnamefont{Ruchowska}},
  \bibinfo{author}{\bibfnamefont{W.~A.} \bibnamefont{Plociennik}},
  \bibinfo{author}{\bibfnamefont{J.}~\bibnamefont{Zylicz}},
  \bibnamefont{et~al.}, \bibinfo{journal}{Phys. Rev. C}
  \textbf{\bibinfo{volume}{73}}, \bibinfo{pages}{044326}
  (\bibinfo{year}{2006}).

\bibitem[{\citenamefont{Rikken and Kessener}(1995)}]{Rikken-95}
\bibinfo{author}{\bibfnamefont{G.~L. J.~A.} \bibnamefont{Rikken}}
  \bibnamefont{and} \bibinfo{author}{\bibfnamefont{Y.~A. R.~R.}
  \bibnamefont{Kessener}}, \bibinfo{journal}{Phys. Rev. Lett.}
  \textbf{\bibinfo{volume}{74}}, \bibinfo{pages}{880} (\bibinfo{year}{1995}).

\bibitem[{\citenamefont{Bodnar and Elster}(2014)}]{Bodnar-14}
\bibinfo{author}{\bibfnamefont{O.}~\bibnamefont{Bodnar}} \bibnamefont{and}
  \bibinfo{author}{\bibfnamefont{C.}~\bibnamefont{Elster}},
  \bibinfo{journal}{Metrologia} \textbf{\bibinfo{volume}{51}},
  \bibinfo{pages}{516} (\bibinfo{year}{2014}).

\bibitem[{\citenamefont{Browne et~al.}(2001)\citenamefont{Browne, Norman,
  Canaan, Glasgow, Keller, and Young}}]{Browne-01}
\bibinfo{author}{\bibfnamefont{E.}~\bibnamefont{Browne}},
  \bibinfo{author}{\bibfnamefont{E.~B.} \bibnamefont{Norman}},
  \bibinfo{author}{\bibfnamefont{R.~D.} \bibnamefont{Canaan}},
  \bibinfo{author}{\bibfnamefont{D.~C.} \bibnamefont{Glasgow}},
  \bibinfo{author}{\bibfnamefont{J.~M.} \bibnamefont{Keller}},
  \bibnamefont{and} \bibinfo{author}{\bibfnamefont{J.~P.} \bibnamefont{Young}},
  \bibinfo{journal}{Phys. Rev. C} \textbf{\bibinfo{volume}{164}},
  \bibinfo{pages}{014311} (\bibinfo{year}{2001}).

\bibitem[{\citenamefont{Kikunaga et~al.}(2009)\citenamefont{Kikunaga,
  Kasamatsu, Haba, Mitsugashira, Hara, Takamiya, Ohtsuki, Yokoyama, Nakanishi,
  and Shinohara}}]{Kikunaga2009}
\bibinfo{author}{\bibfnamefont{H.}~\bibnamefont{Kikunaga}},
  \bibinfo{author}{\bibfnamefont{Y.}~\bibnamefont{Kasamatsu}},
  \bibinfo{author}{\bibfnamefont{H.}~\bibnamefont{Haba}},
  \bibinfo{author}{\bibfnamefont{T.}~\bibnamefont{Mitsugashira}},
  \bibinfo{author}{\bibfnamefont{M.}~\bibnamefont{Hara}},
  \bibinfo{author}{\bibfnamefont{K.}~\bibnamefont{Takamiya}},
  \bibinfo{author}{\bibfnamefont{T.}~\bibnamefont{Ohtsuki}},
  \bibinfo{author}{\bibfnamefont{A.}~\bibnamefont{Yokoyama}},
  \bibinfo{author}{\bibfnamefont{T.}~\bibnamefont{Nakanishi}},
  \bibnamefont{and}
  \bibinfo{author}{\bibfnamefont{A.}~\bibnamefont{Shinohara}},
  \bibinfo{journal}{Phys. Rev. C} \textbf{\bibinfo{volume}{80}},
  \bibinfo{pages}{034315} (\bibinfo{year}{2009}),
  \urlprefix\url{http://link.aps.org/doi/10.1103/PhysRevC.80.034315}.

\bibitem[{\citenamefont{Mitsugashira et~al.}(2003)\citenamefont{Mitsugashira,
  Hara, Ohtsuki, Yuki, Takamiya, Kasamatsu, Shinohara, Kikunaga, and
  Nakanishi}}]{Mitsugashira2003}
\bibinfo{author}{\bibfnamefont{T.}~\bibnamefont{Mitsugashira}},
  \bibinfo{author}{\bibfnamefont{M.}~\bibnamefont{Hara}},
  \bibinfo{author}{\bibfnamefont{T.}~\bibnamefont{Ohtsuki}},
  \bibinfo{author}{\bibfnamefont{H.}~\bibnamefont{Yuki}},
  \bibinfo{author}{\bibfnamefont{K.}~\bibnamefont{Takamiya}},
  \bibinfo{author}{\bibfnamefont{Y.}~\bibnamefont{Kasamatsu}},
  \bibinfo{author}{\bibfnamefont{A.}~\bibnamefont{Shinohara}},
  \bibinfo{author}{\bibfnamefont{H.}~\bibnamefont{Kikunaga}}, \bibnamefont{and}
  \bibinfo{author}{\bibfnamefont{T.}~\bibnamefont{Nakanishi}},
  \bibinfo{journal}{J. Radioanal. Nucl. Chem.} \textbf{\bibinfo{volume}{255}},
  \bibinfo{pages}{63} (\bibinfo{year}{2003}), ISSN \bibinfo{issn}{0236-5731},
  \urlprefix\url{http://dx.doi.org/10.1023/A%3A1022267428310}.

\bibitem[{\citenamefont{Zimmermann}(2010)}]{Zimmermann2010}
\bibinfo{author}{\bibfnamefont{K.}~\bibnamefont{Zimmermann}},
  \bibinfo{type}{Dr. rer. nat.}, \bibinfo{school}{Gottfried Wilhelm Leibniz
  Universit\"{a}t Hannover} (\bibinfo{year}{2010}),
  \urlprefix\url{http://edok01.tib.uni-hannover.de/edoks/e01dh10/634991264.pdf}.

\bibitem[{\citenamefont{{Swanberg Jr.}}(2012)}]{Swanberg2012}
\bibinfo{author}{\bibfnamefont{E.~L.} \bibnamefont{{Swanberg Jr.}}},
  \bibinfo{type}{Dissertation}, \bibinfo{school}{University of California,
  Berkeley} (\bibinfo{year}{2012}),
  \urlprefix\url{http://digitalassets.lib.berkeley.edu/etd/ucb/text/Swanberg_berkeley_0028E_12982.pdf}.

\bibitem[{\citenamefont{Utter et~al.}(1999)\citenamefont{Utter, Beiersdorfer,
  Barnes, Lougheed, Crespo L\'{o}pez-Urrutia, Becker, and Weiss}}]{Utter1999}
\bibinfo{author}{\bibfnamefont{S.~B.} \bibnamefont{Utter}},
  \bibinfo{author}{\bibfnamefont{P.}~\bibnamefont{Beiersdorfer}},
  \bibinfo{author}{\bibfnamefont{A.}~\bibnamefont{Barnes}},
  \bibinfo{author}{\bibfnamefont{R.~W.} \bibnamefont{Lougheed}},
  \bibinfo{author}{\bibfnamefont{J.~R.} \bibnamefont{Crespo
  L\'{o}pez-Urrutia}}, \bibinfo{author}{\bibfnamefont{J.~A.}
  \bibnamefont{Becker}}, \bibnamefont{and}
  \bibinfo{author}{\bibfnamefont{M.~S.} \bibnamefont{Weiss}},
  \bibinfo{journal}{Phys. Rev. Lett.} \textbf{\bibinfo{volume}{82}},
  \bibinfo{pages}{505} (\bibinfo{year}{1999}),
  \urlprefix\url{http://link.aps.org/doi/10.1103/PhysRevLett.82.505}.

\bibitem[{\citenamefont{Shaw et~al.}(1999)\citenamefont{Shaw, Young, Cooper,
  and Webb}}]{Shaw1999}
\bibinfo{author}{\bibfnamefont{R.~W.} \bibnamefont{Shaw}},
  \bibinfo{author}{\bibfnamefont{J.~P.} \bibnamefont{Young}},
  \bibinfo{author}{\bibfnamefont{S.~P.} \bibnamefont{Cooper}},
  \bibnamefont{and} \bibinfo{author}{\bibfnamefont{O.~F.} \bibnamefont{Webb}},
  \bibinfo{journal}{Phys. Rev. Lett.} \textbf{\bibinfo{volume}{82}},
  \bibinfo{pages}{1109} (\bibinfo{year}{1999}),
  \urlprefix\url{http://link.aps.org/doi/10.1103/PhysRevLett.82.1109}.

\bibitem[{Sol()}]{Soldatov-79}
\bibinfo{note}{A. A. Soldatov and D. P. Grechukhin, Kourchatov Institute of
  Atomic Energy Report-3174, Moscow, 1979.}

\bibitem[{\citenamefont{Band and Fomichev}(1979)}]{Band-79}
\bibinfo{author}{\bibfnamefont{I.~M.} \bibnamefont{Band}} \bibnamefont{and}
  \bibinfo{author}{\bibfnamefont{V.~I.} \bibnamefont{Fomichev}},
  \bibinfo{journal}{At. Data Nucl. Data Tabl.} \textbf{\bibinfo{volume}{23}},
  \bibinfo{pages}{295} (\bibinfo{year}{1979}).

\bibitem[{\citenamefont{Nilsson}(1955)}]{Nilsson-55}
\bibinfo{author}{\bibfnamefont{S.~G.} \bibnamefont{Nilsson}},
  \bibinfo{journal}{Mat.-fys medd. danske selskab} \textbf{\bibinfo{volume}{Bd.
  29}}, \bibinfo{pages}{n. 16} (\bibinfo{year}{1955}).

\bibitem[{\citenamefont{Church and Weneser}(1956)}]{Church-56}
\bibinfo{author}{\bibfnamefont{E.~L.} \bibnamefont{Church}} \bibnamefont{and}
  \bibinfo{author}{\bibfnamefont{J.}~\bibnamefont{Weneser}},
  \bibinfo{journal}{Phys. Rev.} \textbf{\bibinfo{volume}{104}},
  \bibinfo{pages}{1382} (\bibinfo{year}{1956}).

\bibitem[{Voi()}]{Voikhansky-66}
\bibinfo{note}{M. E. Voikhansky, M. A. Listengarten, and I. M. Band,
  Penetration effects in internal conversion. In: {\it{Internal Conversion
  Processes}}, Ed. by J.H. Hamilton, Academic Press, NY, 1966, p.581.}

\bibitem[{\citenamefont{Tkalya}(1994)}]{Tkalya-94}
\bibinfo{author}{\bibfnamefont{E.~V.} \bibnamefont{Tkalya}},
  \bibinfo{journal}{JETP Lett.} \textbf{\bibinfo{volume}{78}},
  \bibinfo{pages}{239} (\bibinfo{year}{1994}).

\bibitem[{\citenamefont{Reich and Helmer}(1990)}]{Reich-90}
\bibinfo{author}{\bibfnamefont{C.~W.} \bibnamefont{Reich}} \bibnamefont{and}
  \bibinfo{author}{\bibfnamefont{R.~G.} \bibnamefont{Helmer}},
  \bibinfo{journal}{Phys. Rev. Lett.} \textbf{\bibinfo{volume}{64}},
  \bibinfo{pages}{271} (\bibinfo{year}{1990}).

\bibitem[{Moo()}]{Moore-04}
\bibinfo{note}{I. Moore, I. Ahmad, K. Bailey, D. L. Bowers, Z.-T. Lu, T. P.
  O'Connor, and Z. Yin, Argonne National Laboratory, Physics Division Report
  No. PHY-10990-ME-2004, 2004},
  \urlprefix\url{http://www.phy.anl.gov/mep/atta/publications/thorium229_phy-10990-me-2004.pdf.}

\bibitem[{\citenamefont{Casten}(2000)}]{Casten-2000}
\bibinfo{author}{\bibfnamefont{R.}~\bibnamefont{Casten}},
  \emph{\bibinfo{title}{Nuclear Physics from a Simple Perspective}}
  (\bibinfo{publisher}{Oxford University Press}, \bibinfo{year}{2000}).

\bibitem[{\citenamefont{Tret'yakov et~al.}(1960)\citenamefont{Tret'yakov,
  Anikina, Gol'din, Novikova, and Pirogova}}]{Tretyakov-60}
\bibinfo{author}{\bibfnamefont{E.~F.} \bibnamefont{Tret'yakov}},
  \bibinfo{author}{\bibfnamefont{M.~P.} \bibnamefont{Anikina}},
  \bibinfo{author}{\bibfnamefont{L.~L.} \bibnamefont{Gol'din}},
  \bibinfo{author}{\bibfnamefont{G.~I.} \bibnamefont{Novikova}},
  \bibnamefont{and} \bibinfo{author}{\bibfnamefont{N.~I.}
  \bibnamefont{Pirogova}}, \bibinfo{journal}{Sov. Phys. JETP}
  \textbf{\bibinfo{volume}{37}}, \bibinfo{pages}{656} (\bibinfo{year}{1960}).

\bibitem[{\citenamefont{Kasamatsu et~al.}(2005)\citenamefont{Kasamatsu,
  Kikunaga, Nakashima, Takamiya, Mitsugashira, Nakanishi, Ohtsuki, Yuki, Sato,
  and Shinohara}}]{Kasamatsu2005}
\bibinfo{author}{\bibfnamefont{Y.}~\bibnamefont{Kasamatsu}},
  \bibinfo{author}{\bibfnamefont{H.}~\bibnamefont{Kikunaga}},
  \bibinfo{author}{\bibfnamefont{K.}~\bibnamefont{Nakashima}},
  \bibinfo{author}{\bibfnamefont{K.}~\bibnamefont{Takamiya}},
  \bibinfo{author}{\bibfnamefont{T.}~\bibnamefont{Mitsugashira}},
  \bibinfo{author}{\bibfnamefont{T.}~\bibnamefont{Nakanishi}},
  \bibinfo{author}{\bibfnamefont{T.}~\bibnamefont{Ohtsuki}},
  \bibinfo{author}{\bibfnamefont{H.}~\bibnamefont{Yuki}},
  \bibinfo{author}{\bibfnamefont{W.}~\bibnamefont{Sato}}, \bibnamefont{and}
  \bibinfo{author}{\bibfnamefont{A.}~\bibnamefont{Shinohara}},
  \bibinfo{type}{Research Report} \bibinfo{number}{vol. 38, p. 32--35},
  \bibinfo{institution}{Laboratory of Nuclear Science, Tohoku University}
  (\bibinfo{year}{2005}),
  \urlprefix\url{http://www.lns.tohoku.ac.jp/fy2011/research/report/2005/2005.htm}.

\bibitem[{\citenamefont{Feldman and Cousins}(1998)}]{Feldman1998}
\bibinfo{author}{\bibfnamefont{G.~J.} \bibnamefont{Feldman}} \bibnamefont{and}
  \bibinfo{author}{\bibfnamefont{R.~D.} \bibnamefont{Cousins}},
  \bibinfo{journal}{Phys. Rev. D} \textbf{\bibinfo{volume}{57}},
  \bibinfo{pages}{3873} (\bibinfo{year}{1998}),
  \urlprefix\url{http://link.aps.org/doi/10.1103/PhysRevD.57.3873}.

\bibitem[{\citenamefont{Varshalovich et~al.}(1988)\citenamefont{Varshalovich,
  Moskalev, and Khersonslii}}]{Varshalovich-88}
\bibinfo{author}{\bibfnamefont{D.~A.} \bibnamefont{Varshalovich}},
  \bibinfo{author}{\bibfnamefont{A.~N.} \bibnamefont{Moskalev}},
  \bibnamefont{and} \bibinfo{author}{\bibfnamefont{V.~K.}
  \bibnamefont{Khersonslii}}, \emph{\bibinfo{title}{Quantum Theory of Angular
  Momentum}} (\bibinfo{publisher}{World Scientific Publ.},
  \bibinfo{address}{London}, \bibinfo{year}{1988}).

\end{thebibliography}

\end{document}